%% file: aaskaii_template.tex
\documentclass[a4paper,11pt]{article}
\usepackage{aaskaiid}
\usepackage{orcidlink}
\title{Tracing the Star Formation History of the Universe through Thermal Free-Free Emission with the SKA}
\ShortTitle{Free-free emission with the SKA}

\author[1]{Hiddo S. B. Algera\orcidlink{0000-0002-4205-9567}}
\ShortName{H.S.B. Algera et al.} 
\author[2]{Mark Sargent\orcidlink{0000-0003-1033-9684}}
\author[3]{Eric J.\ Murphy}
\author[4]{Fangxia An\orcidlink{0000-0001-7943-0166}}

\affiliation[1]{Institute of Astronomy and Astrophysics, Academia Sinica, 11F of Astronomy-Mathematics Building, No.1, Sec. 4, Roosevelt Rd, Taipei 106319, Taiwan, R.O.C.}
\emailAdd{hsbalgera@asiaa.sinica.tw.edu}
\affiliation[2]{Institute of Physics, Laboratory of Astrophysics, \'Ecole Polytechnique F\'ed\'erale de Lausanne (EPFL), Observatoire de Sauverny, Versoix CH-1290, Switzerland}
\affiliation[3]{National Radio Astronomy Observatory, 520 Edgemont Road, Charlottesville, VA 22903, USA}
\affiliation[4]{Yunnan Observatories, Chinese Academy of Sciences, Kunming 650216, People’s Republic of China}

\abstract{One of the major scientific aims of the SKA is to trace the history of star formation across cosmic time. High-frequency radio surveys are indispensable in this regard, as these are capable of probing thermal free-free emission (FFE) -- the dominant component of the radio continuum of star-forming galaxies above rest-frame frequencies of $\gtrsim25\,$GHz. FFE is a powerful, direct star-formation rate (SFR) indicator, which robustly traces the number of ionizing photons produced by recently formed massive stars in a nearly dust-unbiased manner. In this chapter, we forecast the ability of the SKA to detect FFE in typical star-forming galaxies in the early Universe. Our starting point is the state-of-the-art T-RECS simulation suite of the faint radio sky, to which we apply an ambitious, matched-depth, multi-band AA4 SKA-Mid survey in Bands 1 through 5b, covering an area of $0.25\,\mathrm{deg}^2$ across $0.35 - 15.4\,\mathrm{GHz}$. We predict that such a survey will detect $\sim1.5\times10^4$ star-forming galaxies in all bands out to $z\approx7$, and perform simulations using established fitting techniques to investigate the accuracy with which their thermal FFE can be recovered. We find that thermal fractions ($f_\mathrm{th}$) and synchrotron spectral indices can be constrained in an unbiased manner, and predict uncertainties on the thermal SFRs of $\lesssim 0.1\,\mathrm{dex}$ for galaxies at the knee of the radio luminosity function across redshift. Convolving the distribution of $f_\mathrm{th}$ inferred from the multi-band SKA-Mid survey with wider luminosity function determinations at low radio frequencies will yield robust constraints on the total cosmic star formation rate density out to $z\sim7$.
}

\begin{document}
\maketitle

\section{Introduction}
One of the key aims in extragalactic astronomy is to trace the star formation rate density (SFRD) across cosmic time \citep[e.g.,][]{madau_dickinson2014}. While optical and near-infrared facilities such as the \textit{Hubble}, \textit{Spitzer} and \textit{James Webb} space telescopes have mapped the contribution of unobscured star formation to well beyond $z\gtrsim10$ \citep[e.g.,][]{mclure2013,bouwens2015,oesch2018,finkelstein2023}, these facilities cannot readily account for star formation hidden behind dust. Such obscured star formation is known to dominate the SFRD out to at least $z\lesssim4$ \citep{madau_dickinson2014,zavala2021}, and possibly well beyond \citep{gruppioni2020,algera2023,sun2025}. While current (sub-)millimeter facilities such as ALMA are sensitive enough to detect obscured star formation in individual galaxies well beyond Cosmic Noon ($z\sim2-3$, coincident with the peak of the cosmic SFRD), as demonstrated by recent high-redshift detections \citep[e.g.,][]{hashimoto2019, fudamoto2021, inami2022}, they lack the mapping speed to carry out large, blind surveys at these early epochs. Consequently, our understanding of obscured star formation and thus the build-up of galaxies across time remains severely limited beyond $z\gtrsim3$ \citep[c.f.,][]{casey2018}.

Radio continuum emission provides a clear way forward in mapping the high-redshift SFRD \citep[e.g.,][]{novak2017,leslie2020,vandervlugt2022}. At GHz frequencies, the radio emission from star-forming galaxies tightly correlates with their star formation rates (SFRs) through what is known as the infrared/radio correlation \citep[IRRC; e.g.,][]{condon1992,bell2003}. This is due to the infrared emission from these galaxies -- which accounts for the bulk of their SFR -- being powered by the same young, massive stars that drive their non-thermal synchrotron emission once they have ended their lives as core-collapse supernovae. However, the exact nature of the IRRC in the early Universe, and whether it evolves with redshift or galaxy properties such as stellar mass, remains poorly understood \citep[e.g.,][]{sargent2010a,sargent2010b,delhaize2017,algera2020_irrc,delvecchio2021,dezotti2024}. Consequently, SFRs derived through the IRRC currently remain systematically uncertain. While SKA-Mid is set to greatly advance our understanding of the IRRC at early cosmic times (see the companion chapter in this volume by \citealt{FangxiaAn01.2026.SKA}), it is paramount to simultaneously explore other, more direct SFR tracers in the early Universe.

Fortunately, the radio spectrum provides such a tracer. Thermal free-free emission (FFE) -- which becomes the dominant emission mechanism at rest-frame $\nu \gtrsim 25\,\mathrm{GHz}$ \citep{condon1992,murphy2011} -- provides a direct ($\lesssim10\,\mathrm{Myr}$) and nearly completely dust-unbiased proxy of galaxy SFRs, which have lauded it as a particularly powerful tracer in the early Universe \citep[e.g.,][]{murphy2017,jimenez-andrade2024}. However, its intrinsic faintness and high-frequency nature have limited direct observations of FFE in the early Universe to just a handful of galaxies \citep[e.g.,][]{thomson2012,huynh2017,algera2021,chen2024}. The high sensitivity and mapping speed provided by SKA-Mid are set to change this, as discussed in detail in this Chapter.

\begin{figure}
    \centering
    \includegraphics[width=0.9\textwidth]{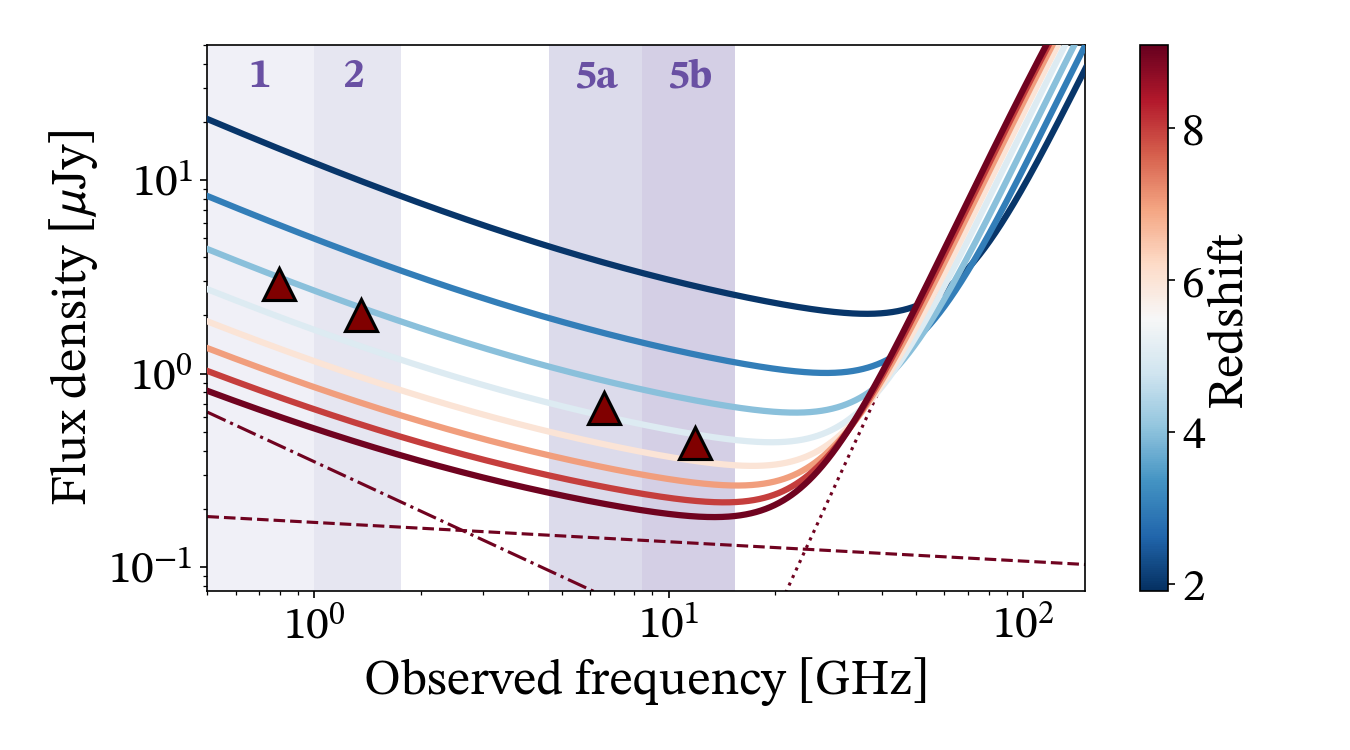}
    \caption{The typical long-wavelength spectrum of a star-forming galaxy with $\mathrm{SFR} = 100\,M_\odot\,\mathrm{yr}^{-1}$, plotted for a range of redshifts between $z=2 - 9$. The radio component assumes a thermal fraction of $f_\mathrm{th}(1.4\,\mathrm{GHz}) = 0.1$ and a synchrotron slope of $\alpha_\mathrm{NT} = -0.85$, while dust emission is modeled as a modified blackbody with temperature $T_\mathrm{dust} = 35\,\mathrm{K}$ and emissivity index $\beta_\mathrm{IR}=2.0$. The individual free-free, synchrotron and dust components of the $z=9$ spectrum (bottom) are indicated through the dash-dotted, dashed and dotted lines, respectively. The nominal frequency coverage of SKA-Mid Bands 1, 2, 5a and 5b is indicated through the vertical shading, and the maroon triangles represent the $5\sigma$ detection limits for the fiducial SKA-Mid survey outlined in Section \ref{sec:survey}, which is set to directly detect highly star-forming galaxies ($\mathrm{SFR} \gtrsim 100\,M_\odot\,\mathrm{yr}^{-1}$) in all bands well beyond Cosmic Noon ($z\gtrsim3$).}
    \label{fig:exampleSED}
\end{figure}

The typical radio spectrum of a star-forming galaxy with $\mathrm{SFR}=100\,M_\odot\,\mathrm{yr}^{-1}$, redshifted between $z = 2 - 9$, is shown in Figure \ref{fig:exampleSED}. The radio emission of star-forming galaxies is often presumed to be composed of two components \citep[e.g.,][]{condon1992,niklas1997,murphy2017,tabatabaei2017,tabatabaei2025,klein2018,algera2021}: non-thermal synchrotron emission with a relatively steep spectral slope ($\alpha_\mathrm{NT} \sim -0.8$), and flat-spectrum thermal free-free emission ($\alpha_\mathrm{FF} = -0.1$).\footnote{We define the spectral index as $\alpha = d\log S_\nu / d\log\nu$, i.e., we use the sign convention whereby $S_\nu \propto \nu^{\alpha}$ such that typically $\alpha < 0$. Throughout this chapter we add the subscript `NT' (i.e., $\alpha_\mathrm{NT}$) when referring specifically to the non-thermal synchrotron spectral index.} At higher frequencies ($\nu \gtrsim 200\,\mathrm{GHz}$ in the galaxy rest-frame), thermal dust emission is expected to overtake the radio emission. This picture is somewhat of an oversimplification, as the radio spectra of high-redshift SFGs can also exhibit spectral curvature \citep[e.g.,][]{calistro-rivera2017,galvin2018,thomson2019,tisanic2019,an2021}. Moreover, anomalous microwave emission (AME) is known to contribute to radio emission at higher frequencies \citep[e.g.,][]{murphy2010,murphy2018}. These contributions to the radio spectrum for low-redshift galaxies are discussed in the chapter by \citet{Moldon01.2026.SKA} in this volume, which we refer to for details.

For simplicity, however, throughout this Chapter we will assume a two-component radio spectrum where synchrotron and free-free emission are both represented by single power-laws. Since we focus on studying FFE at high redshifts, the effects of (sub-)GHz-frequency spectral turnovers and flattening are limited for our parameter space of interest (i.e., SKA-Mid Band 1 already probes $\nu_\mathrm{rest} \gtrsim 3\,\mathrm{GHz}$ beyond $z\gtrsim3$). Moreover, the main population that SKA-Mid will be able to detect consists of faint, star-forming galaxies typically beyond the confusion limit in the SKA-Low bands, where a detailed low-frequency characterization will thus generally not be feasible. Nevertheless, capturing these complex radio spectra for bright high-redshift galaxies is well within the reach of SKA-Low and -Mid, and will be key to better understand, for instance, injection processes of cosmic ray electrons. Fully characterizing spectral curvature and other higher-order effects with SKA-Mid would furthermore greatly benefit from expanding its frequency coverage to Bands 3 and 4, and ideally even beyond its Band 5 \citep[see also the entry by M.\ Sargent et al.\ in the SKA Memo by][]{conway2018}.

\section{A Multi-band SKA-Mid Survey of Radio Free-free emission}
\label{sec:survey}

The primary goal of this Chapter is to forecast the power of SKA-Mid AA4 to reliably trace thermal free-free emission in distant galaxies. To investigate this, we first adopt a realistic multi-frequency SKA-Mid survey, following the strategy outlined in the companion chapter in this volume by \citet{Prandoni01.2026.SKA} which is summarized below. Subsequently, we apply these survey parameters to the \citet{bonaldi2019} simulations of the radio sky to generate a realistic population of mock radio sources, akin to those that will be observed by SKA-Mid. Finally, we decompose their spectra into synchrotron and free-free emission following established fitting approaches. We detail this process below.

\subsection{A SKA-Mid AA4 multi-frequency survey}
\label{ref:surveyParameters}

To assess the ability of SKA-Mid to probe FFE in faint, star-forming galaxies, our starting point is the deep, multi-frequency SKA survey outlined in the companion chapter by \citet{Prandoni01.2026.SKA} in this volume, which in turn builds upon the SKA-Mid Deep survey outlined in \citet{prandoni_seymour2015}. In summary, their proposed survey is a $0.25\,\mathrm{deg}^2$, matched-depth SKA-Low + SKA-Mid survey in all available observing bands, i.e., spanning a range of $80\,\mathrm{MHz} - 11.85\,\mathrm{GHz}$. The proposed RMS noise in Band 2 ($1.355\,\mathrm{GHz}$) is approximately $0.4\,\mu\mathrm{Jy/beam}$, which is then scaled to all other bands assuming a standard spectral index of $\alpha = -0.70$. In Band 5b, this corresponds to a noise level of $0.09\,\mu\mathrm{Jy/beam}$. In Bands 1 and 2, this requires an observing time of approximately $10\,\mathrm{h}$ each, while mosaicking is required to cover this field-of-view at higher frequencies. As a result, Bands 5a and 5b dominate the total SKA-Mid time request, with indicative observing times of $\sim500\,\mathrm{h}$ (12 pointings) and $\sim3300\,\mathrm{h}$ (39 pointings), respectively.

As outlined in \citet[][their Section 3.2.2]{Prandoni01.2026.SKA}, the proposed survey aims for an approximately matched-depth, matched-resolution view of the high-redshift galaxy population, and adopts $uv$-tapering and different imaging weighting schemes to do so. For the SKA-Mid Bands, this results in a typical angular resolution of $\sim2'' - 2.5''$; ideal for robust flux density measurements of faint radio sources, as these are not expected to be resolved on these scales \citep[e.g.,][]{miettinen2017,jimenez-andrade2019}.

Although this proposed multi-frequency survey probes all the way down to SKA-Low frequencies, we here focus only on the full suite of SKA-Mid Bands (i.e., Bands 1, 2, 5a and 5b). This is because the diagnostic power of low-frequency measurements in tracing free-free emission is limited given the typical deviation from a two-component nature in this regime \citep{calistro-rivera2017,galvin2018,thomson2019}. Overall, we thus consider a frequency range of $\nu_\mathrm{obs} = 0.80 - 12\,\mathrm{GHz}$ in our analysis, which corresponds to the central frequencies of SKA-Mid Bands 1 through 5b, as also indicated in Figure \ref{fig:exampleSED}.

\subsection{Generating a realistic SKA-Mid galaxy population}

We apply the survey parameters detailed above to the \citet{bonaldi2019} T-RECS simulations, which simulate the extragalactic radio sky from $150\,\mathrm{MHz} - 20\,\mathrm{GHz}$. Both star-forming galaxies (SFGs), radio-quiet and radio-loud active galactic nuclei (AGN) are included in the simulations, although \citet{bonaldi2019} model the radio-quiet AGN as a subset of star-forming galaxies. We here follow this approach, and do not further distinguish between these populations, though we briefly comment on how such AGN may impact studies of radio free-free emission in Section \ref{sec:discussion_caveats}. Moreover, we will not consider radio-loud AGN in this Chapter, and thus essentially assume that these can be robustly identified through ancillary multi-frequency observations. In practice this is not always a straightforward task, although radio-loud AGN make up only a small fraction of the faint radio sky such that their expected level of contamination is limited in deep SKA-Mid surveys \citep[e.g.,][]{delvecchio2017,smolcic2017_multiwave,algera2020_xs,lyu2022,peluso2025}.

\begin{figure}
    \centering
    \includegraphics[width=0.9\textwidth]{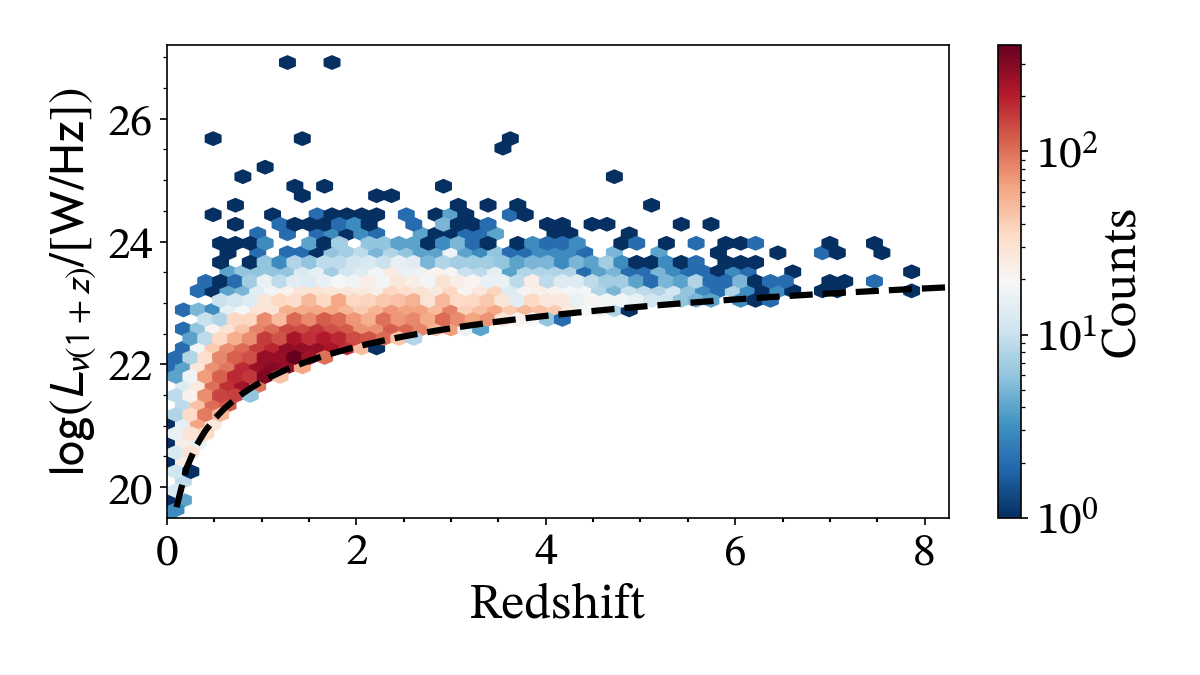}
    \caption{Radio luminosity at the rest-frame frequency probed by SKA-Mid Band 2 ($\nu_\mathrm{obs}=1.355\,\mathrm{GHz}$) against redshift for galaxies sampled in the central $0.25\,\mathrm{deg}^2$ of an SKA-Mid Band 2 pointing with a $5\sigma = 2.0\,\mu\mathrm{Jy/beam}$ depth. The luminosity limit of the survey is indicated through the dashed black line. Approximately $1.5\times10^4$ radio sources -- the vast majority ($\gtrsim95\%$) of which are star-forming galaxies -- are expected to be detected above this limit, of which $\sim2000$ reside at $z>3$ and $\sim50$ at $z>6$.}
    \label{fig:sensitivityCurve}
    \vspace*{-0.3cm}
\end{figure}

We extract mock star-forming galaxies within a single, randomized SKA-Mid Band 2 field-of-view from the \citet{bonaldi2019} simulations at a reference frequency of $\nu_\mathrm{cen} = 1.4\,\mathrm{GHz}$.\footnote{The catalogs provided by \citet{bonaldi2019} contain flux densities at a fixed set of frequencies; 1.4$\,$GHz is closest to the central frequency of $1.355\,\mathrm{GHz}$ adopted in the \citet{Prandoni01.2026.SKA} survey. We do not correct for the minor difference in frequency, which affects flux density measurements to $\lesssim2 - 3\,\%$ given a typical $\alpha = -0.70$.} We subsequently apply a $5\sigma$ point-source detection limit, folding in attenuation by the primary beam, which we approximate as a Gaussian with $\mathrm{FWHM} = 48.3'$. The Band 2 sensitivity limit of our mock survey as well as the full set of sampled mock sources (SFGs + AGN) from the T-RECS simulations is shown in Figure \ref{fig:sensitivityCurve}. The expected number of radio sources detectable at $>5\sigma$ in the central $0.25\,\mathrm{deg}^2$ of the Band 2 pointing is approximately $N\sim1.5\times10^4$, of which $\gtrsim95\%$ are expected to be star-forming galaxies (or radio-quiet AGN). By virtue of the matched-depth nature of the radio survey adopted in this Chapter, the majority of these are detected at $>5\sigma$ across all four SKA-Mid bands.

\subsection{Constructing the mock radio spectra}

While the T-RECS simulations readily provide multi-frequency flux densities for the simulated sources, we here use the SKA-Mid Band 2 flux density as a starting point from which to construct their mock radio spectra. For each of the mock sources, we draw a thermal fraction normalized at rest-frame $1.4\,\mathrm{GHz}$, denoted $f_\mathrm{th}(1.4\,\mathrm{GHz})$, from a normal distribution with mean $\mu=0.10$ and standard deviation $\sigma = 0.06$, clipped between $f_\mathrm{th}(1.4\,\mathrm{GHz})\in[0.01, 0.30]$. We furthermore draw a synchrotron spectral index $\alpha_\mathrm{NT}$ from a normal distribution with $\mu=-0.85$ and $\sigma=0.30$, clipped between $\alpha_\mathrm{NT}\in[-1.5, -0.3]$. These parameters are broadly representative for star-forming galaxies at high redshift \citep[e.g.,][]{murphy2017,smolcic2017_release,algera2021,an2021}. Combined, this fully determines the shape of their radio spectra, which we subsequently sample in the four SKA-Mid bands. We fold in simple Gaussian noise based on the RMS of the images, including the effects of the primary beam in Bands 1 and 2. For Bands 5a and 5b, where mosaicking is required, we assume no primary beam attenuation. The assumption of purely Gaussian noise effectively implies the galaxies are fully unresolved at the adopted resolution of $2-2.5''$, such that the uncertainty on their (peak) flux density is well-represented by the local noise \citep[see also][]{condon1997}.

\begin{figure}
    \centering
    \includegraphics[width=0.8\textwidth]{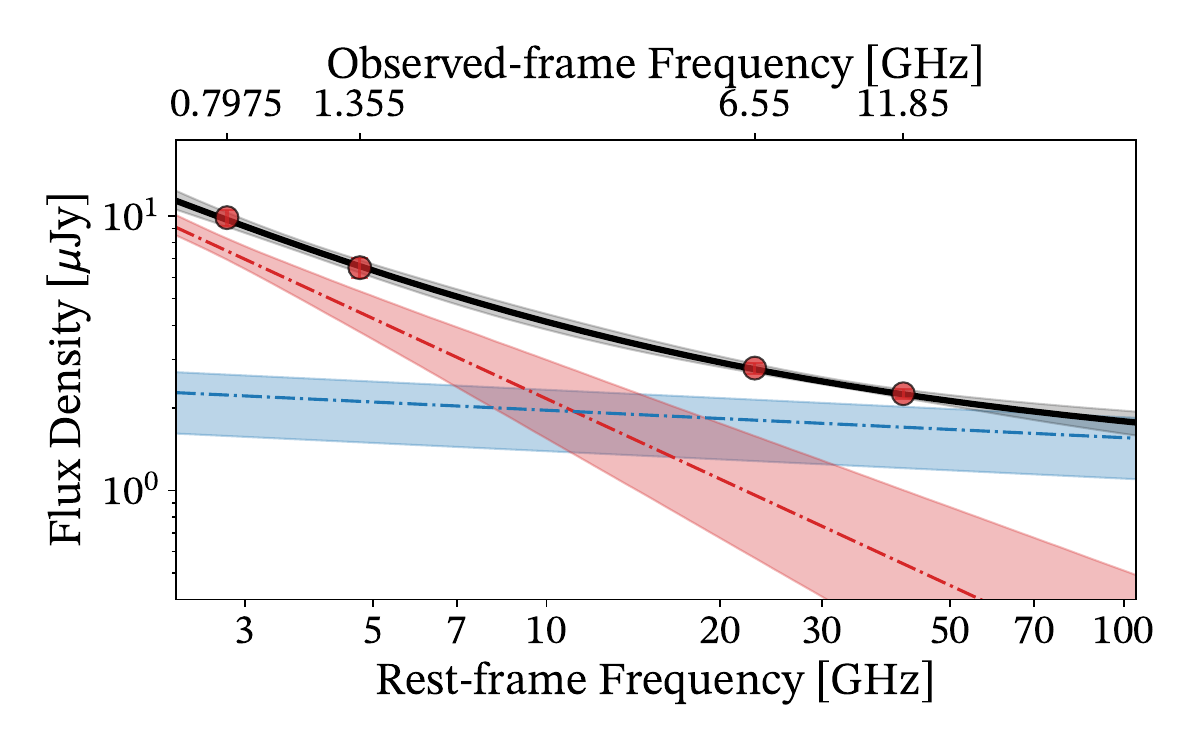}
    \caption{The radio spectrum of an example mock $z\approx2.5$ galaxy, as will be identified in large numbers in upcoming deep SKA-Mid surveys. The radio spectrum is sampled in Bands 1, 2, 5a and 5b (red datapoints; central observed-frame frequencies are annotated on the top $x$-axis) and decomposed into its synchrotron (red) and free-free (blue) components. The black line and grey shading represent the total spectrum and its uncertainty, respectively. The signal-to-noise in the four bands spans $\mathrm{S/N} = 12 - 26$, which enables the accurate detection of a flattening spectrum at high frequencies; the signature of thermal free-free emission.}
    \label{fig:spectralDecompositionExample}
\end{figure}

\subsection{Decomposing the mock radio spectra}
\label{sec:spectralDecomposition}

The next step is decomposing the mock radio spectra into their synchrotron and free-free components. We adopt the mock flux density measurements in Bands 1, 2, 5a and 5b, as outlined in the previous section, and fit these using the Bayesian framework from \citet{algera2021,algera2022}. Briefly, this approach uses the Monte Carlo Markov Chain (MCMC) framework implemented in the {\sc{emcee}} library \citep{foreman-mackey2013}, and fits for three free parameters: the flux density at an arbitrary frequency, here chosen to be $1.4\,\mathrm{GHz}$ in the observer-frame (denoted $S_{1.4}$); the thermal fraction at this frequency; and the synchrotron spectral index. A flat prior on the thermal fraction is adopted, between $f_\mathrm{th}\in[-0.2, 1.2]$. Unphysical values below zero and above unity are allowed as these yield more reliable uncertainties and can be used to assess the need for a thermal component in the fit \citep{linden2020,algera2021}. We furthermore adopt a Gaussian prior on the synchrotron slope with mean $\mu = -0.85$ and standard deviation $\sigma = 0.5$, as in \citet{algera2021} and \citet{chen2024}, and a weakly informative, flat prior on the flux density $S_{1.4}$. An example of the decomposed radio spectrum for a mock galaxy at $z=2.5$ is shown in Figure \ref{fig:spectralDecompositionExample}.

\section{Predicted constraints on high-redshift free-free emission with SKA-Mid}
\label{sec:results}

\subsection{Recovery of thermal fractions and synchrotron slopes}
\label{sec:results_recovery}

\begin{figure}
    \centering
    \includegraphics[width=0.495\textwidth]{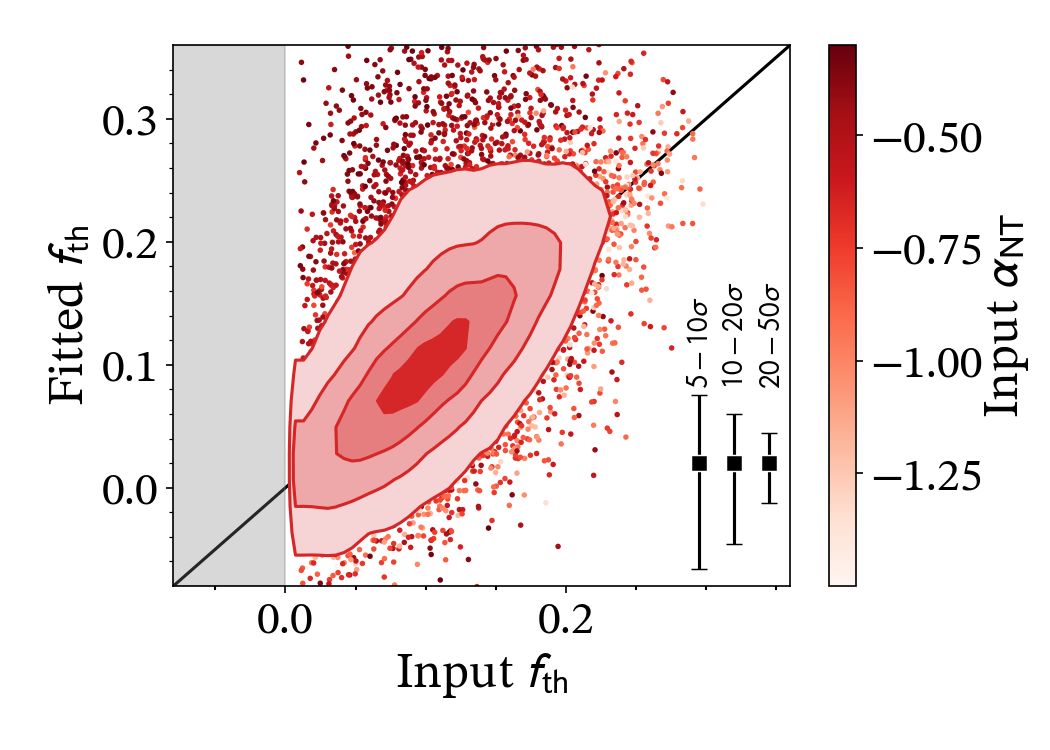}
    \includegraphics[width=0.495\textwidth]{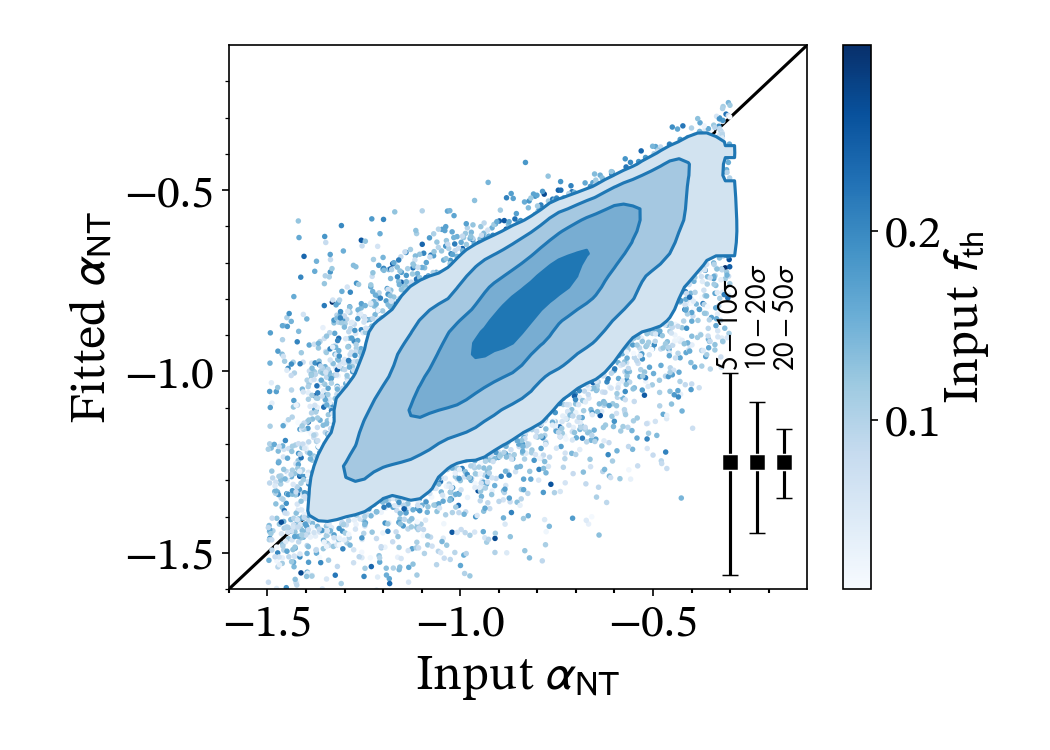}
    \caption{The fitted thermal fractions (\textit{left}) and synchrotron spectral indices (\textit{right}) of our mock radio sources as a function of their input values. Contours are drawn at the $0.5, 1, 1.5$ and $2\sigma$ levels, and individual datapoints are color-coded by $\alpha_\mathrm{NT}$ (\textit{left}) or $f_\mathrm{th}(1.4\,\mathrm{GHz})$ (\textit{right}). Three representative errorbars are shown in the bottom right corner of each panel, and correspond to different bins in Band 2 S/N ($5-10\sigma$, $10-20\sigma$, and $20-50\sigma$). In both panels, the input and fitted values agree well with one another, suggesting that the radio spectra of star-forming galaxies can be well-recovered in our simulated multi-band SKA-Mid survey. Only a minority population of galaxies with flat synchrotron slopes (dark red points in the \textit{left} panel) can induce some bias in measurements of the thermal fraction.}
    \label{fig:spectralDecompositionResults}
\end{figure}

We perform the spectral decomposition outlined in the previous section on the full sample of $\sim1.5\times10^4$ mock star-forming galaxies in our fiducial SKA-Mid survey. The fitted thermal fractions and synchrotron slopes are shown as a function of their input values in Figure \ref{fig:spectralDecompositionResults}. Overall, we find good agreement between the fitted and input parameters, suggesting that our simulated SKA-Mid survey can aptly recover the intrinsic radio spectra of star-forming galaxies, and robustly isolate their free-free emission. However, some caveats apply: the left panel of Figure \ref{fig:spectralDecompositionResults} reveals a population of outliers with high fitted thermal fractions ($f_\mathrm{th}(1.4\,\mathrm{GHz}) \gtrsim0.3$). These are all characterized by a shallow input synchrotron slope, highlighting the known degeneracy between a large free-free contribution and an overall flat synchrotron spectrum \citep[c.f.,][]{algera2021,chen2024}. In the right panel of the Figure, the correlation between the fitted and input synchrotron slopes is slightly shallower than unity. This is due to the adopted Gaussian prior on $\alpha_\mathrm{NT}$ in the modeling, which down-weights fits with particularly shallow or steep synchrotron spectra, and thus slightly flattens the relation between the fitted and input slopes.

\begin{figure}
    \centering
    \includegraphics[width=0.6\textwidth]{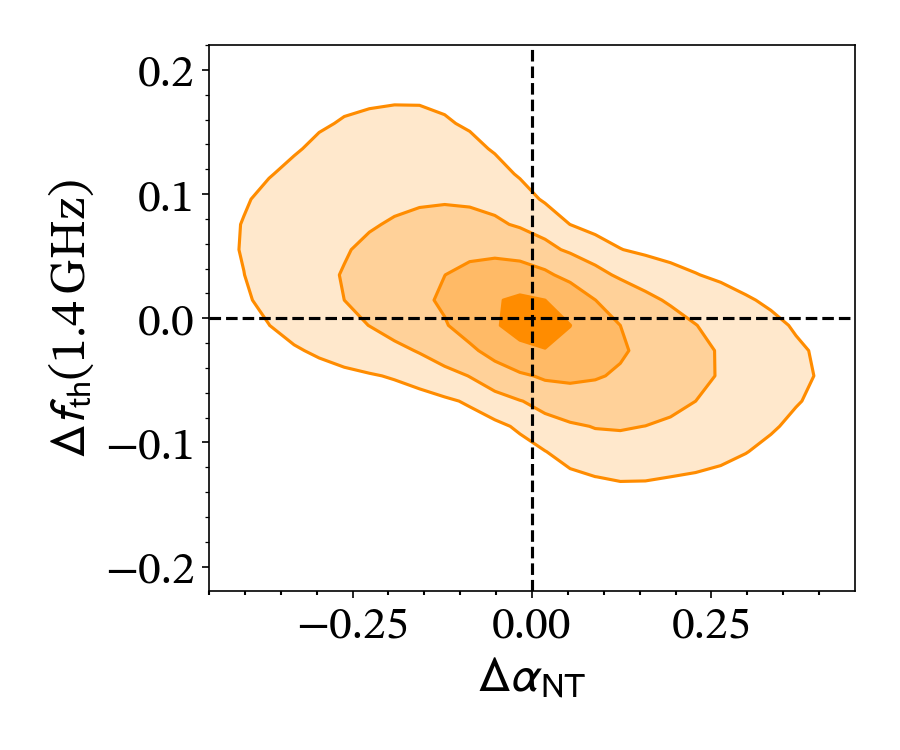}
    \caption{Offset between fitted and input thermal fractions (at 1.4$\,$GHz; $\Delta f_\mathrm{th} = f_\mathrm{th,fit} - f_\mathrm{th,input}$) versus that for the synchrotron slope ($\Delta\alpha_\mathrm{NT}$, defined analogously) across the full population of mock sources. There are no systematic biases in the recovery of either parameter, although a mild anti-correlation is visible.}
    \label{fig:spectralDecompositionFthVsAnt}
\end{figure}

Figure \ref{fig:spectralDecompositionFthVsAnt} plots the difference between the fitted and input parameters ($\Delta f_\mathrm{th} = f_\mathrm{th,fit} - f_\mathrm{th,input}$ and similar for $\Delta\alpha_\mathrm{NT}$) against each other. The two parameters are mildly anti-correlated, but across the full mock sample of $\sim1.5\times10^4$ star-forming galaxies neither shows any systematic biases. We infer a median $\Delta f_\mathrm{th} = 0.00_{-0.04}^{+0.06}$ and $\Delta\alpha_\mathrm{NT} = -0.01_{-0.19}^{+0.14}$ across the full sample (here, and elsewhere in this Chapter, the errors represent the $16-84^\mathrm{th}$ percentile spread of the distribution).

\begin{figure}
    \centering
    \includegraphics[width=0.495\textwidth]{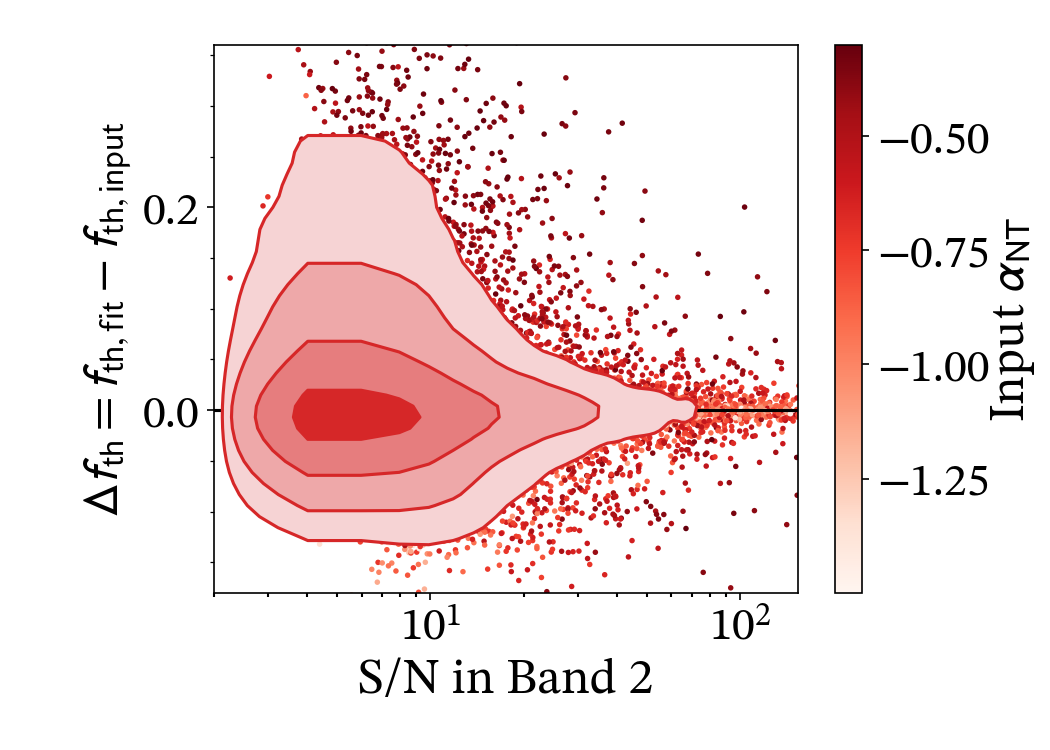}
    \includegraphics[width=0.495\textwidth]{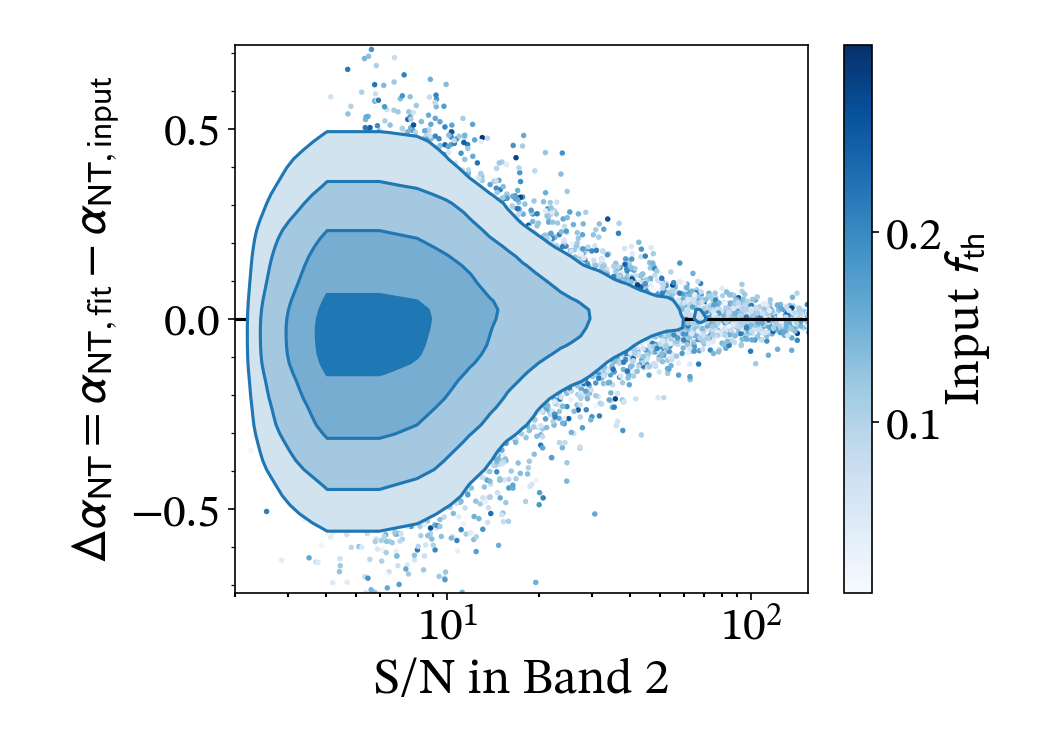}
    \caption{The accuracy with which the thermal fractions (\textit{left}) and synchrotron spectral indices (\textit{right}) can be recovered, as a function of the Band 2 signal-to-noise ratio. Contours and points are as in Figure \ref{fig:spectralDecompositionResults}. Deep SKA-Mid surveys will not be systematically biased in their measurements of thermal free-free emission even at low S/N, although uncertainties will be significant in this regime. This is mainly due to the degeneracy between $f_\mathrm{th}$ and $\alpha_\mathrm{NT}$ that requires high S/N ($\gtrsim10 - 20$) or observations in additional bands to break. The accuracy with which the synchrotron slope can be recovered is a strong function of Band 2 S/N, as expected, but similarly does not show any systematic biases at low signal-to-noise. }
    \label{fig:spectralDecompositionResultsSNR}
\end{figure}

While the above discussion establishes the overall accuracy with which the spectra can be decomposed into their thermal and non-thermal components, the recovery of the fitting parameters is expected to depend on the signal-to-noise ratio (S/N) in the radio maps. We show $\Delta f_\mathrm{th}(1.4\,\mathrm{GHz})$ and $\Delta \alpha_\mathrm{NT}$ as a function of the S/N in SKA-Mid Band 2 -- our nominal `detection band' in this exercise -- in Figure \ref{fig:spectralDecompositionResultsSNR}.

The synchrotron spectral index is well-recovered across the full range of S/N expected from the T-RECS simulations, without any systematic biases or offsets. The scatter in $\alpha_\mathrm{NT}$ unsurprisingly increases towards lower S/N, and extends out to the width of the Gaussian prior near the detection limit ($0.5\,\mathrm{dex}$; Section \ref{sec:spectralDecomposition}). The thermal fraction is similarly not strongly biased, although the aforementioned population with shallow synchrotron slopes yields an uptick in $\Delta f_\mathrm{th}$ at low signal-to-noise due to fitting degeneracies between the two parameters. Nevertheless, we conclude that, on average, the spectral decomposition is robust even at modest S/N ($\lesssim10$).

\begin{figure}
    \centering
    \includegraphics[width=0.495\textwidth]{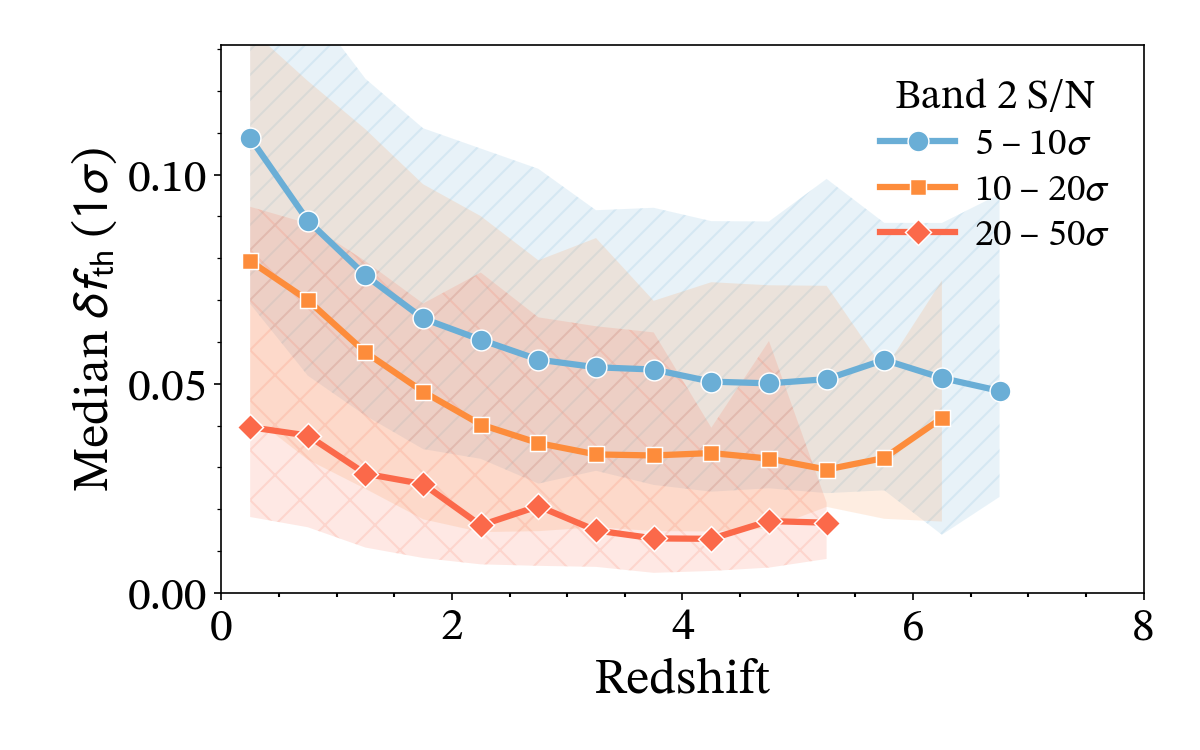}
    \includegraphics[width=0.495\textwidth]{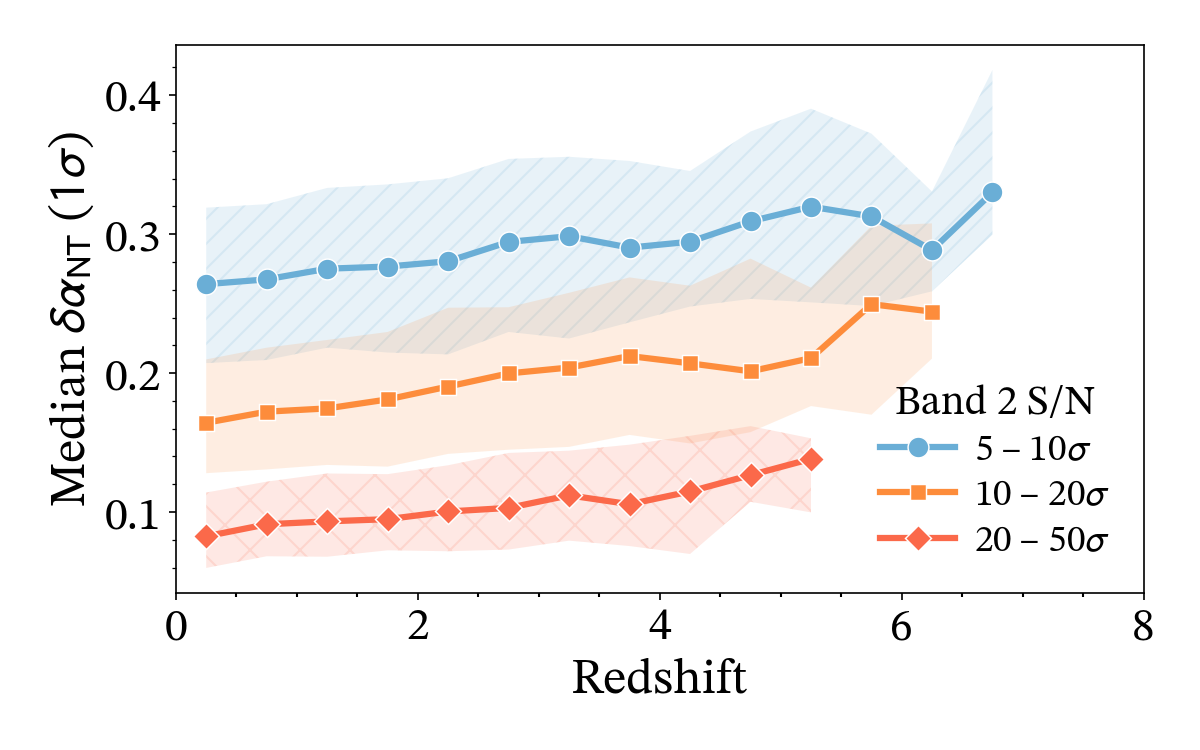}
    \caption{The typical $1\sigma$ uncertainty on the thermal fraction (\textit{left}) and synchrotron spectral index (\textit{right}) as a function of redshift. Three bins of SKA-Mid Band 2 S/N are considered, and only those with $\geq5$ sources are plotted. The solid lines represent the median uncertainty on $f_\mathrm{th}(1.4\,\mathrm{GHz})$ and $\alpha_\mathrm{NT}$ in a given bin, while the shaded regions represent the $16-84^\mathrm{th}$ percentile scatter. At a fixed signal-to-noise ratio, thermal fractions can be better constrained at higher redshift, while the accuracy with which the synchrotron slope can be measured decreases slightly towards earlier cosmic times.}
    \label{fig:uncertaintiesVsRedshiftAndSNR}
\end{figure}

Given that we are ultimately interested in tracing the evolution of the galaxy population across time, we finally consider the recovery of the thermal fraction and synchrotron slope as a function of redshift. We construct three signal-to-noise bins in SKA-Mid Band 2 ($5-10\sigma$, $10-20\sigma$, and $20 - 50\sigma$) and compute the median uncertainty on $f_\mathrm{th}(1.4\,\mathrm{GHz})$ and $\alpha_\mathrm{NT}$, denoted $\delta f_\mathrm{th}$ and $\delta \alpha_\mathrm{NT}$, in bins of redshift with a width of $\Delta z = 0.5$. Only bins with $N \geq 5$ sources are considered.

The resulting parameter uncertainties are shown in Figure \ref{fig:uncertaintiesVsRedshiftAndSNR}. Higher-S/N bins naturally yield tighter constraints on both the thermal fraction and synchrotron slope. More notably, at fixed S/N, the thermal fraction is typically better constrained at higher redshifts, given that the SKA-Mid bands better trace the free-free-dominated regime at these earlier epochs. Conversely, the accuracy with which $\alpha_\mathrm{NT}$ can be measured decreases slightly towards higher redshifts. In the next Section, we discuss how these uncertainties propagate into the expected constraints on free-free star formation rates for our fiducial SKA-Mid survey.

\subsection{Constraints on thermal SFRs}
\label{sec:resultsThermalSFR}

The quantity that we are ultimately interested in constraining is the free-free star formation rate, $\mathrm{SFR}_\mathrm{FF}$. Under the assumption of particular initial mass function (IMF) and electron temperature $T_e$, this can be determined as $\mathrm{SFR}_\mathrm{FF} \propto f_\mathrm{th}(\nu) L_\nu = L_\nu^\mathrm{th}$ \citep[e.g.,][]{murphy2011}, where $L_\nu$ is the radio luminosity at rest-frame frequency $\nu$ and $L_\nu^\mathrm{th}$ the corresponding thermal free-free luminosity. Both the thermal fraction and total radio luminosity can be determined at any arbitrary frequency by propagating (a representative sample of draws from) the MCMC posterior distributions, and we here adopt a canonical frequency of $1.4\,\mathrm{GHz}$. We define the `free-free S/N' as $\mathrm{S/N}_\mathrm{FF} = L_{1.4}^\mathrm{th} / \delta L_{1.4}^\mathrm{th}$, i.e., the ratio of the thermal luminosity and its (symmetrized) uncertainty, which is equal to the S/N on the thermal SFR given the above assumptions.

\begin{figure}
    \centering
    \includegraphics[width=0.9\textwidth]{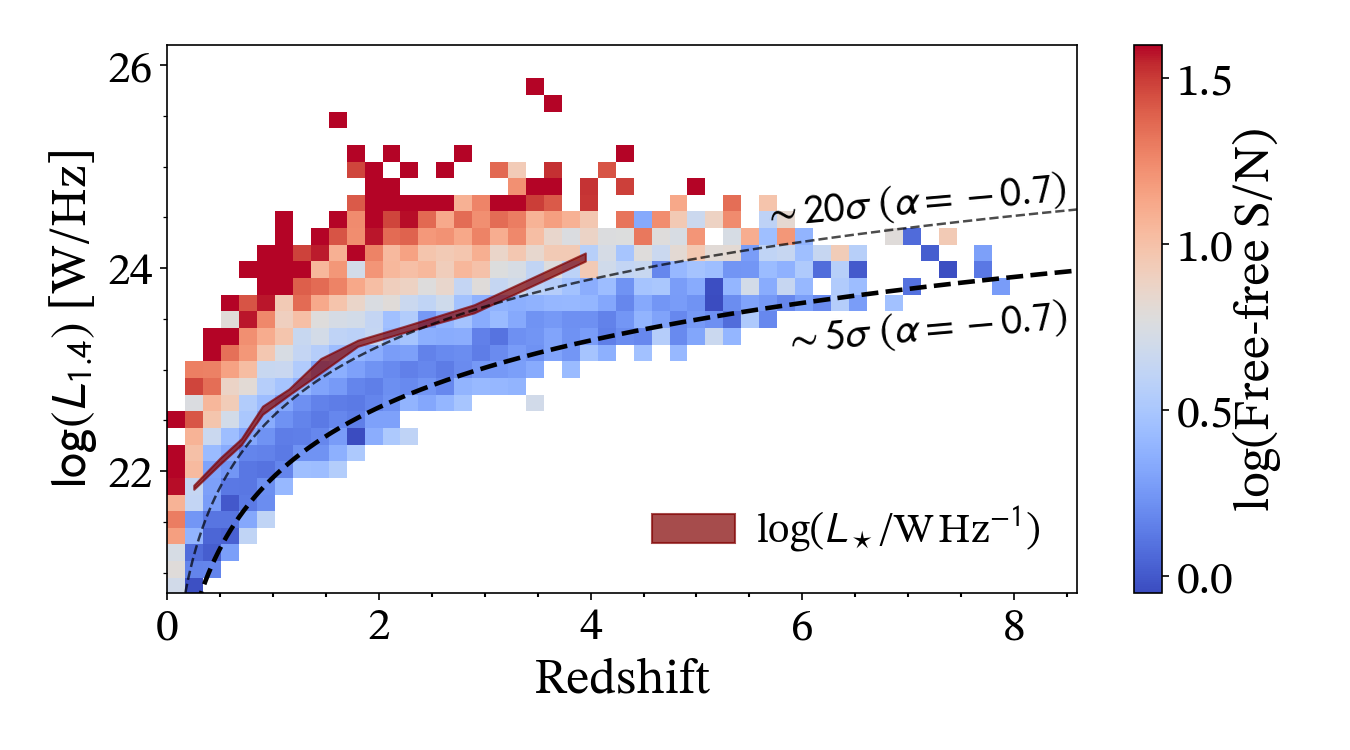} 
    \caption{Binned rest-frame $1.4\,\mathrm{GHz}$ radio luminosity versus redshift for our mock SKA-Mid survey. The bins are color-coded by their mean free-free S/N. The thick dashed line corresponds to the approximate sensitivity limit, based on the $5\sigma$ SKA-Mid Band 2 detection threshold of $2\,\mu\mathrm{Jy/beam}$ and a fixed $\alpha = -0.70$. The thin line represents sources that are $4\times$ brighter than this limit (i.e., detected at the $\sim20\sigma$ level). The maroon shading corresponds to the knee of the luminosity function $L_\star$ determined by \citet{cochrane2023} at $150\,\mathrm{MHz}$ and scaled to $1.4\,\mathrm{GHz}$ with $\alpha=-0.7$. Free-free emission in galaxies at the knee can be readily detected (at $>5-10\sigma$) out to $z\sim6$.}
    \label{fig:sensitivityCurveSNR}
\end{figure}

We show the mean $\mathrm{S/N}_\mathrm{FF}$ in bins of redshift and radio luminosity in Figure \ref{fig:sensitivityCurveSNR}. Near the detection limit ($\sim5\sigma$), the uncertainties on free-free SFRs will be $\gtrsim50\,\%$, although for bright SFGs the S/N rapidly increases.\footnote{In our simplified model, $\mathrm{S/N}_\mathrm{FF} > 50$ is readily achievable in particularly bright sources. However, at this stage second-order effects in the shape of the radio spectrum will dominate the error budget.} The maroon shading in Figure \ref{fig:sensitivityCurveSNR} shows the `knee' of the radio luminosity function (LF) from \citet{cochrane2023}, who employed sensitive LOFAR observations across a $\sim25\,\mathrm{deg}^2$ area to constrain the LF at $150\,\mathrm{MHz}$ out to $z\sim4$. We scale their results to $1.4\,\mathrm{GHz}$ assuming a standard $\alpha=-0.70$, and find that galaxies at the knee of the LF will be detected at approximately $\sim20\sigma$ in our fiducial SKA-Mid survey.\footnote{We note that \citet{calistro-rivera2017} indeed find a typical spectral index of $\alpha^{150}_{1400}\approx-0.70$ for SFGs such that this scaling is appropriate. Moreover, we obtain consistent results when adopting the $3\,\mathrm{GHz}$-based measurements of $L_\star$ from \citet{vandervlugt2022}, which are based on deeper data, albeit across a smaller area.} At this S/N, the FFE component can be readily constrained at $>5\sigma$ in individual galaxies, implying the nominal uncertainty on their thermal SFRs will be $\lesssim20\,\%$ (i.e., $\lesssim0.1\,\mathrm{dex}$). We expect such robust constraints for $\sim1000$ galaxies beyond $z>2$, of which $\sim150$ lie beyond $z>4$. 

To cast the knee of the $1.4\,\mathrm{GHz}$ LF into a more directly interpretable quantity -- namely star formation rate -- we make use of the infrared/radio correlation. This provides an SFR measurement mostly independent of free-free emission based on the predominantly non-thermal radio continuum emission at lower frequencies. At a typical redshift of $z = 3$, where the radio LF remains well-constrained observationally, $L_\star$ corresponds to $\mathrm{SFR}_\mathrm{1.4\,GHz}\sim75 - 100\,M_\odot\,\mathrm{yr}^{-1}$ assuming the redshift-dependent infrared/radio correlation of \citet{delhaize2017} or the redshift-invariant but non-linear correlation of \citet{molnar2021}. This, in turn, roughly corresponds to the typical star-formation rate of a main-sequence galaxy with stellar mass $M_\star \approx 10^{10.5}\,M_\odot$ at $z\approx3$ \citep{speagle2014}.

We next investigate to what extent $\mathrm{S/N}_\mathrm{FF}$ depends on redshift for our realistic population of radio-detected star-forming galaxies. At fixed radio luminosity, more distant galaxies are fainter due to the positive $K$-correction, though in Section \ref{sec:results_recovery} we also predicted that -- at fixed S/N -- constraints on FFE improve at higher redshifts. We focus only on the mock sample with $L_{1.4} > 10^{24}\,\mathrm{W/Hz}$, which is expected to be detectable at $>5\sigma$ regardless of redshift. For this highly star-forming subset, the median free-free S/N in individual galaxies is expected to decrease from $\mathrm{S/N}_\mathrm{FF}\sim30$ at $1\lesssim z \lesssim 2$ to $\mathrm{S/N}_\mathrm{FF}\sim7$ at $4 \lesssim z \lesssim 5$. This thus suggests that, at fixed radio luminosity, the ability of SKA-Mid to constrain FFE will decrease with increasing redshift. Nevertheless, for sufficiently luminous galaxies, free-free emission will be readily detectable out to $z\sim6$.

\subsection{Constraints on cosmic star formation}
\label{sec:results_SFRD}

In the previous sections, we have demonstrated that -- for a reasonable input distribution of $f_\mathrm{th}$ and a realistic population of radio-detected SFGs -- it is possible to recover the true thermal fraction with high accuracy (Figures \ref{fig:spectralDecompositionResults} through \ref{fig:uncertaintiesVsRedshiftAndSNR}). This will enable the determination of the `free-free luminosity function' $\Phi(L_\mathrm{th})$ as a function of redshift, analogously to the standard radio luminosity function. While necessitating detailed simulations to account for completeness, integrating $\Phi(L_\mathrm{th})$ will provide a direct constraint on the SFRD through radio free-free emission. 

An arguably more powerful means of constraining the SFRD, however, will be to calibrate the predominantly non-thermal low-frequency radio emission as a star formation rate tracer using free-free-derived SFRs as a benchmark. While FFE and synchrotron emission trace star formation on different timescales \citep[e.g.,][]{bressan2002}, in practice such variation will mostly average out when considering large galaxy samples. As a result, the prime objective will be to find the calibration function $g(p)$ such that $\mathrm{SFR}_\mathrm{FF} = g(p) \times L_{1.4}$, where $p=\{z, M_\star, \ldots\}$ is the possible parameter set upon which the conversion from low-frequency radio luminosity to star formation rate depends. Various techniques, such as Bayesian model comparison or a Principal Component Analysis (PCA), can then be used to quantify which physical parameters are the main drivers of variation in the $L_{1.4}$-to-SFR conversion. Once an appropriate functional form has been established, it will be possible to accurately use low-frequency radio observations -- where survey speeds are significantly higher than in SKA-Mid Bands 5a and 5b -- to probe the SFRD well into the epoch of reionization ($z\gtrsim6$). 

Equivalently, the SFRD can be determined by convolving the distribution of $f_\mathrm{th}(1.4\,\mathrm{GHz})$ obtained from our fiducial multi-band SKA-Mid survey with the $1.4\,\mathrm{GHz}$ luminosity function, which can be readily measured across much wider areas. This can be expressed as

\begin{equation}
    \mathrm{SFRD}(z) = C(T_e, \mathrm{IMF}) \int \Phi(L_{1.4}) f_\mathrm{th}(1.4\,\mathrm{GHz}; L_{1.4}, z, \ldots) d \log L_{1.4}
\end{equation}

where the proportionality factor $C$ depends on the electron temperature and IMF, which will need to be assumed a priori, or marginalized over. The key functional form to constrain in this approach will be the dependence of the thermal fraction on $L_{1.4}$, as well as any possible additional dependencies on -- for instance -- stellar mass.

\subsection{Challenges and multi-wavelength synergies}
\label{sec:discussion_caveats}

\subsubsection{Redshifts}

Throughout this Chapter, we have highlighted the power of free-free emission as a star-formation rate tracer out to high redshift ($z\sim6$). In doing so, we have implicitly assumed that the redshifts of the radio-detected galaxy population can be robustly constrained, which is required for accurate radio luminosity measurements, as well as for a robust spectral decomposition. In practice, obtaining redshifts requires multi-band optical-to-near-infrared (NIR) photometry or -- ideally -- spectroscopy, and the availability of such data would therefore naturally greatly benefit the proposed multi-band SKA-Mid survey.

While the SKA will efficiently survey the southern and equatorial radio sky at low frequencies, studies of free-free emission will remain limited to relatively smaller areas due to the required high-frequency capabilities and therefore limited instantaneous field-of-view. As a result, a deep $\sim0.25\,\mathrm{deg}^2$ survey such as that proposed by \citet{Prandoni01.2026.SKA} and further explored in this Chapter is ideally suited to complement existing multi-wavelength deep fields, leveraging the rich photometric and spectroscopic datasets from facilities like \textit{HST}, \textit{JWST}, and ALMA.

As an example, approximately $\sim0.5\,\mathrm{deg}^2$ across the $\sim2\,\mathrm{deg}^2$ COSMOS field has been covered by deep \textit{JWST}/NIRCam imaging \citep{casey2023}, including a $\sim0.2\,\mathrm{deg}^2$ area targeted at mid-infrared wavelengths with MIRI as well as recently at millimeter wavelengths by ALMA as part of the \textit{CHAMPS} program \citep[PI Faisst, see also][]{zavala2026}. The availability of such optical-to-mm data would provide important context in which the multi-frequency radio observations can be interpreted -- not only for (photometric) redshift measurements, but also for obtaining robust stellar masses, and for identifying AGN. Similarly, COSMOS has received a wealth of spectroscopic follow-up \citep[e.g.,][]{khostovan2026}, thereby yielding a high spectroscopic completeness for a possible SKA-Mid survey carried out across this part of the sky. Other southern/equatorial deep fields, such as UDS \citep{lawrence2007}, GOODS-South \citep{giavalisco2004}, and the \textit{Euclid} Deep Field South \citep{scaramella2022} represent similarly promising targets for deep, high-frequency efforts with SKA-Mid.

Finally, we highlight that ongoing and future spectroscopic surveys and facilities are poised to greatly increase the spectroscopic completeness of the distant galaxy population. Facilities and instruments such as Subaru/PFS \citep{tamura2016}, VISTA/4MOST \citep{dejong2019}, and the upcoming VLT/MOONS \citep{cirasuolo2014} combine powerful multiplexing capabilities with a high sensitivity, thereby enabling particularly efficient redshift surveys. In the more distant future, dedicated facilities such as the proposed Wide-field Spectroscopic Telescope \citep[WST;][]{mainieri2024} may further expand such spectroscopic coverage to wider areas and fainter galaxies.

\subsubsection{Deviations from a two-component radio model: biases and physical implications}

In our analysis, we have assumed that the radio spectrum can be well-represented by a combination of two power laws, comprising a steep and flat component representing synchrotron emission and free-free emission, respectively. We here briefly highlight how deviations from this simple model could bias the inferred thermal fractions, and at the same time yield novel physical insights into the galaxies powering the radio emission.

First of all, if the radio spectrum were to systematically flatten at lower frequencies, for example due to free-free absorption (FFA), this would preferentially suppress the SKA-Mid Band 1 emission compared to our fiducial model. This, in turn, would yield shallower synchrotron spectral indices -- which are mostly driven by the low-frequency constraints -- and therefore lower overall thermal fractions and SFRs. While the effects of FFA may be expected to decrease at high redshift given the higher rest-frame frequencies being probed, galaxies are also known to become increasingly compact and dense at earlier times \citep[e.g.,][]{vanderwel2014,isobe2023}, which in turn would increase the relative importance of FFA. Multi-frequency constraints at low frequencies, likely through a combination of SKA-Low and SKA-Mid Bands 1 and 2, are therefore crucial to assess if, and to what extent, such flattening occurs in typical high-redshift SFGs.

At higher radio frequencies, synchrotron aging is moreover known to steepen the radio spectrum \citep[e.g.,][]{thomson2019}. As discussed in \citet{algera2022}, this too could lead to thermal fractions being underestimated, as such steepening can counteract the expected flattening of the radio spectrum through free-free emission. While the necessarily young ages of high-redshift galaxies could on the one hand limit the effects of spectral aging, their possibly stronger magnetic fields would instead serve to increase it \citep[e.g.,][]{schleicher2013,algera2020_irrc,yoon2024}. Similarly, inverse-Compton losses -- possibly against a warmer Cosmic Microwave Background \citep[e.g.,][]{murphy2009,whittam2025} -- would similarly preferentially suppress the high-frequency synchrotron component given its frequency-dependence of $\tau_\mathrm{IC} \propto \nu^{-1/2}$. A detailed characterization of spectral aging with SKA-Mid will not be straightforward given the limited high-frequency coverage, but could plausibly be detected through a systematic steepening between Bands 5a and 5b. This can furthermore be explored as a function of redshift, and thereby provide insights into -- for example -- the evolution of magnetic fields in the galaxy population across time.

Finally, we have assumed throughout this chapter that radio AGN can be accurately separated from the SFG sample on which the spectral decomposition is carried out. For radio-loud AGN, this is a reasonable assumption as these can be readily identified through their deviation from the infrared/radio correlation \citep[e.g.,][]{yun2001,delmoro2013,smolcic2017_multiwave} by making use of ancillary far-infrared data from facilities such as \textit{Herschel}, ALMA and, in the near future, \textit{PRIMA} \citep{glenn2025}.

However, low-level radio emission from faint AGN cannot easily be identified in this manner, and could therefore potentially boost the observed synchrotron emission from a galaxy classified as star-forming, and thus reduce the inferred thermal fraction. AGN signatures at other wavelengths, such as from X-rays or rest-optical emission line ratios \citep[e.g.,][]{bonzini2013,delvecchio2017}, can be employed to define a `clean' SFG sample, although the extent of which such sources also show AGN signatures at radio wavelengths remains debated \citep[e.g.,][]{padovani2011,herrera-ruiz2017,radcliffe2021}. An additional powerful means of identifying faint radio AGN is to leverage the high native angular resolution of the high-frequency SKA-Mid bands. While our fiducial multi-wavelength survey employs tapering to obtain a matched $\sim2.5''$ resolution across all bands, SKA-Mid Band 5b can attain a resolution of just a few milliarcseconds through standard Briggs weighting. This enables further AGN identification through the detection of bright, compact sources as well as small-scale radio jets \citep[e.g.,][]{herrera-ruiz2017,radcliffe2018,morabito2025}. While (faint) radio AGN could thus ostensibly be a source of contamination, the combination of high-resolution radio observations and powerful ancillary optical through far-infrared data will help to partially mitigate this.

\section{Summary}

In this Chapter, we have presented predictions on the prospects of detecting free-free emission from star-forming galaxies in the early Universe using the baseline AA4 design of the SKA. Our starting point is the SKA-Mid component of the multi-band radio survey proposed by \citet{Prandoni01.2026.SKA}, who advocate for matched-depth, matched-resolution observations in Bands 1, 2, 5a and 5b. Using the T-RECS simulations \citep{bonaldi2019}, we extract a realistic population of $\sim1.5\times10^4$ star-forming galaxies across a $\sim0.25\,\mathrm{deg}^2$ field-of-view, to which we apply the proposed Band 2 detection limit ($5\sigma\sim2\,\mu\mathrm{Jy/beam}$). Adopting reasonable distributions for the thermal fraction and synchrotron spectral index, we determine the expected flux densities of this sample in the other SKA-Mid bands in the presence of Gaussian noise. Through established fitting techniques \citep{algera2021,algera2022}, we decompose the radio spectra into their free-free and synchrotron components and forecast the accuracy with which the intrinsic distributions of $f_\mathrm{th}$ and $\alpha_\mathrm{NT}$ can be recovered. 

We find that both the thermal fraction and synchrotron slope can be recovered in an unbiased manner, meaning there are no systematic offsets between the input and fitted parameter values. As expected, the significance with which the free-free component can be extracted depends primarily on the detection S/N. For galaxies detected at modest significance ($\sim5\sigma$ per band), uncertainties on free-free SFRs will be $\gtrsim50\,\%$. However, for galaxies detected at $\gtrsim20\sigma$, which corresponds to galaxies roughly at the knee of the radio luminosity function, the error will be $\lesssim20\,\%$, implying uncertainties on thermal SFRs of $\lesssim0.1\,\mathrm{dex}$. Such robust constraints will be common out to Cosmic Noon, and remain feasible for particularly luminous star-forming galaxies out to $z\sim6$. 

Armed with a sample of radio sources in which free-free emission is robustly detected, it will be possible to determine the SFRD out to $z\sim6$ directly by constructing and subsequently integrating the free-free radio luminosity function. A more powerful approach, however, will be to calibrate the conversion between low-frequency radio emission (e.g., at the canonical $1.4\,\mathrm{GHz}$) and free-free star-formation rate, which should represent the `ground truth' measurement. Given the increased survey speed of SKA-Mid at lower frequencies, this will enable determining the SFRD from much larger galaxy samples. 

While we caveat that we have made some simplifying assumptions -- such as the approximation that the radio spectrum at $\gtrsim\mathrm{GHz}$ frequencies can be described by two power laws, and that AGN contamination is negligible -- our analysis demonstrates the powerful ability of SKA-Mid to trace radio free-free emission well beyond Cosmic Noon, and into the epoch of reionization.

\bibliographystyle{abbrvnat-maxbibnames4}
\bibliography{chapter}

\end{document}

%% file: chapter.bib
@ARTICLE{madau_dickinson2014,
       author = {{Madau}, Piero and {Dickinson}, Mark},
        title = "{Cosmic Star-Formation History}",
      journal = {\araa},
         year = 2014,
        month = aug,
       volume = {52},
        pages = {415-486},
          doi = {10.1146/annurev-astro-081811-125615},
archivePrefix = {arXiv},
       eprint = {1403.0007},
 primaryClass = {astro-ph.CO},
       adsurl = {https://ui.adsabs.harvard.edu/abs/2014ARA&A..52..415M}
}

@ARTICLE{zavala2021,
       author = {{Zavala}, J.~A. and {Casey}, C.~M. and {Manning}, S.~M. and {Aravena}, M. and {Bethermin}, M. and {Caputi}, K.~I. and {Clements}, D.~L. and {Cunha}, E. da and {Drew}, P. and {Finkelstein}, S.~L. and {Fujimoto}, S. and {Hayward}, C. and {Hodge}, J. and {Kartaltepe}, J.~S. and {Knudsen}, K. and {Koekemoer}, A.~M. and {Long}, A.~S. and {Magdis}, G.~E. and {Man}, A.~W.~S. and {Popping}, G. and {Sanders}, D. and {Scoville}, N. and {Sheth}, K. and {Staguhn}, J. and {Toft}, S. and {Treister}, E. and {Vieira}, J.~D. and {Yun}, M.~S.},
        title = "{The Evolution of the IR Luminosity Function and Dust-obscured Star Formation over the Past 13 Billion Years}",
      journal = {\apj},
         year = 2021,
        month = mar,
       volume = {909},
       number = {2},
          eid = {165},
        pages = {165},
          doi = {10.3847/1538-4357/abdb27},
archivePrefix = {arXiv},
       eprint = {2101.04734},
 primaryClass = {astro-ph.GA},
       adsurl = {https://ui.adsabs.harvard.edu/abs/2021ApJ...909..165Z}
}

@ARTICLE{sun2025,
       author = {{Sun}, Fengwu and {Wang}, Feige and {Yang}, Jinyi and {Champagne}, Jaclyn B. and {Decarli}, Roberto and {Fan}, Xiaohui and {Ba{\~n}ados}, Eduardo and {Cai}, Zheng and {Colina}, Luis and {Egami}, Eiichi and {Hennawi}, Joseph F. and {Jin}, Xiangyu and {Jun}, Hyunsung D. and {Khusanova}, Yana and {Li}, Mingyu and {Li}, Zihao and {Lin}, Xiaojing and {Liu}, Weizhe and {Meyer}, Romain A. and {Pudoka}, Maria A. and {Rieke}, George H. and {Shen}, Yue and {Tee}, Wei Leong and {Venemans}, Bram and {Walter}, Fabian and {Wu}, Yunjing and {Zhang}, Huanian and {Zou}, Siwei},
        title = "{A SPectroscopic Survey of Biased Halos in the Reionization Era (ASPIRE): Spectroscopically Complete Census of Obscured Cosmic Star Formation Rate Density at z = 4{\textendash}6}",
      journal = {\apj},
         year = 2025,
        month = feb,
       volume = {980},
       number = {1},
          eid = {12},
        pages = {12},
          doi = {10.3847/1538-4357/ad9d0e},
archivePrefix = {arXiv},
       eprint = {2412.06894},
 primaryClass = {astro-ph.GA},
       adsurl = {https://ui.adsabs.harvard.edu/abs/2025ApJ...980...12S}
}

@ARTICLE{algera2023,
       author = {{Algera}, Hiddo S.~B. and {Inami}, Hanae and {Oesch}, Pascal A. and {Sommovigo}, Laura and {Bouwens}, Rychard J. and {Topping}, Michael W. and {Schouws}, Sander and {Stefanon}, Mauro and {Stark}, Daniel P. and {Aravena}, Manuel and {Barrufet}, Laia and {da Cunha}, Elisabete and {Dayal}, Pratika and {Endsley}, Ryan and {Ferrara}, Andrea and {Fudamoto}, Yoshinobu and {Gonzalez}, Valentino and {Graziani}, Luca and {Hodge}, Jacqueline A. and {Hygate}, Alexander P.~S. and {de Looze}, Ilse and {Nanayakkara}, Themiya and {Schneider}, Raffaella and {van der Werf}, Paul P.},
        title = "{The ALMA REBELS survey: the dust-obscured cosmic star formation rate density at redshift 7}",
      journal = {\mnras},
         year = 2023,
        month = feb,
       volume = {518},
       number = {4},
        pages = {6142-6157},
          doi = {10.1093/mnras/stac3195},
archivePrefix = {arXiv},
       eprint = {2208.08243},
 primaryClass = {astro-ph.GA},
       adsurl = {https://ui.adsabs.harvard.edu/abs/2023MNRAS.518.6142A}
}

@ARTICLE{gruppioni2020,
       author = {{Gruppioni}, C. and {B{\'e}thermin}, M. and {Loiacono}, F. and {Le F{\`e}vre}, O. and {Capak}, P. and {Cassata}, P. and {Faisst}, A.~L. and {Schaerer}, D. and {Silverman}, J. and {Yan}, L. and {Bardelli}, S. and {Boquien}, M. and {Carraro}, R. and {Cimatti}, A. and {Dessauges-Zavadsky}, M. and {Ginolfi}, M. and {Fujimoto}, S. and {Hathi}, N.~P. and {Jones}, G.~C. and {Khusanova}, Y. and {Koekemoer}, A.~M. and {Lagache}, G. and {Lemaux}, B.~C. and {Oesch}, P.~A. and {Pozzi}, F. and {Riechers}, D.~A. and {Rodighiero}, G. and {Romano}, M. and {Talia}, M. and {Vallini}, L. and {Vergani}, D. and {Zamorani}, G. and {Zucca}, E.},
        title = "{The ALPINE-ALMA [CII] survey. The nature, luminosity function, and star formation history of dusty galaxies up to z ≃ 6}",
      journal = {\aap},
         year = 2020,
        month = nov,
       volume = {643},
          eid = {A8},
        pages = {A8},
          doi = {10.1051/0004-6361/202038487},
archivePrefix = {arXiv},
       eprint = {2006.04974},
 primaryClass = {astro-ph.GA},
       adsurl = {https://ui.adsabs.harvard.edu/abs/2020A&A...643A...8G}
}

@ARTICLE{hashimoto2019,
       author = {{Hashimoto}, Takuya and {Inoue}, Akio K. and {Mawatari}, Ken and {Tamura}, Yoichi and {Matsuo}, Hiroshi and {Furusawa}, Hisanori and {Harikane}, Yuichi and {Shibuya}, Takatoshi and {Knudsen}, Kirsten K. and {Kohno}, Kotaro and {Ono}, Yoshiaki and {Zackrisson}, Erik and {Okamoto}, Takashi and {Kashikawa}, Nobunari and {Oesch}, Pascal A. and {Ouchi}, Masami and {Ota}, Kazuaki and {Shimizu}, Ikkoh and {Taniguchi}, Yoshiaki and {Umehata}, Hideki and {Watson}, Darach},
        title = "{Big Three Dragons: A z = 7.15 Lyman-break galaxy detected in [O III] 88 {\ensuremath{\mu}}m, [C II] 158 {\ensuremath{\mu}}m, and dust continuum with ALMA}",
      journal = {\pasj},
         year = 2019,
        month = aug,
       volume = {71},
       number = {4},
          eid = {71},
        pages = {71},
          doi = {10.1093/pasj/psz049},
archivePrefix = {arXiv},
       eprint = {1806.00486},
 primaryClass = {astro-ph.GA},
       adsurl = {https://ui.adsabs.harvard.edu/abs/2019PASJ...71...71H}
}

@ARTICLE{inami2022,
       author = {{Inami}, Hanae and {Algera}, Hiddo S.~B. and {Schouws}, Sander and {Sommovigo}, Laura and {Bouwens}, Rychard and {Smit}, Renske and {Stefanon}, Mauro and {Bowler}, Rebecca A.~A. and {Endsley}, Ryan and {Ferrara}, Andrea and {Oesch}, Pascal and {Stark}, Daniel and {Aravena}, Manuel and {Barrufet}, Laia and {da Cunha}, Elisabete and {Dayal}, Pratika and {De Looze}, Ilse and {Fudamoto}, Yoshinobu and {Gonzalez}, Valentino and {Graziani}, Luca and {Hodge}, Jacqueline A. and {Hygate}, Alexander P.~S. and {Nanayakkara}, Themiya and {Pallottini}, Andrea and {Riechers}, Dominik A. and {Schneider}, Raffaella and {Topping}, Michael and {van der Werf}, Paul},
        title = "{The ALMA REBELS Survey: dust continuum detections at z > 6.5}",
      journal = {\mnras},
         year = 2022,
        month = sep,
       volume = {515},
       number = {3},
        pages = {3126-3143},
          doi = {10.1093/mnras/stac1779},
archivePrefix = {arXiv},
       eprint = {2203.15136},
 primaryClass = {astro-ph.GA},
       adsurl = {https://ui.adsabs.harvard.edu/abs/2022MNRAS.515.3126I}
}

@ARTICLE{delvecchio2021,
       author = {{Delvecchio}, I. and {Daddi}, E. and {Sargent}, M.~T. and {Jarvis}, M.~J. and {Elbaz}, D. and {Jin}, S. and {Liu}, D. and {Whittam}, I.~H. and {Algera}, H. and {Carraro}, R. and {D'Eugenio}, C. and {Delhaize}, J. and {Kalita}, B.~S. and {Leslie}, S. and {Moln{\'a}r}, D. Cs. and {Novak}, M. and {Prandoni}, I. and {Smol{\v{c}}i{\'c}}, V. and {Ao}, Y. and {Aravena}, M. and {Bournaud}, F. and {Collier}, J.~D. and {Randriamampandry}, S.~M. and {Randriamanakoto}, Z. and {Rodighiero}, G. and {Schober}, J. and {White}, S.~V. and {Zamorani}, G.},
        title = "{The infrared-radio correlation of star-forming galaxies is strongly M$_{{\ensuremath{\star}}}$-dependent but nearly redshift-invariant since z {\ensuremath{\sim}} 4}",
      journal = {\aap},
         year = 2021,
        month = mar,
       volume = {647},
          eid = {A123},
        pages = {A123},
          doi = {10.1051/0004-6361/202039647},
archivePrefix = {arXiv},
       eprint = {2010.05510},
 primaryClass = {astro-ph.GA},
       adsurl = {https://ui.adsabs.harvard.edu/abs/2021A&A...647A.123D}
}

@ARTICLE{delhaize2017,
       author = {{Delhaize}, J. and {Smol{\v{c}}i{\'c}}, V. and {Delvecchio}, I. and {Novak}, M. and {Sargent}, M. and {Baran}, N. and {Magnelli}, B. and {Zamorani}, G. and {Schinnerer}, E. and {Murphy}, E.~J. and {Aravena}, M. and {Berta}, S. and {Bondi}, M. and {Capak}, P. and {Carilli}, C. and {Ciliegi}, P. and {Civano}, F. and {Ilbert}, O. and {Karim}, A. and {Laigle}, C. and {Le F{\`e}vre}, O. and {Marchesi}, S. and {McCracken}, H.~J. and {Salvato}, M. and {Seymour}, N. and {Tasca}, L.},
        title = "{The VLA-COSMOS 3 GHz Large Project: The infrared-radio correlation of star-forming galaxies and AGN to z {\ensuremath{\lesssim}} 6}",
      journal = {\aap},
         year = 2017,
        month = jun,
       volume = {602},
          eid = {A4},
        pages = {A4},
          doi = {10.1051/0004-6361/201629430},
archivePrefix = {arXiv},
       eprint = {1703.09723},
 primaryClass = {astro-ph.GA},
       adsurl = {https://ui.adsabs.harvard.edu/abs/2017A&A...602A...4D}
}

@ARTICLE{sargent2010a,
       author = {{Sargent}, M.~T. and {Schinnerer}, E. and {Murphy}, E. and {Aussel}, H. and {Le Floc'h}, E. and {Frayer}, D.~T. and {Mart{\'\i}nez-Sansigre}, A. and {Oesch}, P. and {Salvato}, M. and {Smol{\v{c}}i{\'c}}, V. and {Zamorani}, G. and {Brusa}, M. and {Cappelluti}, N. and {Carilli}, C.~L. and {Carollo}, C.~M. and {Ilbert}, O. and {Kartaltepe}, J. and {Koekemoer}, A.~M. and {Lilly}, S.~J. and {Sanders}, D.~B. and {Scoville}, N.~Z.},
        title = "{The VLA-COSMOS Perspective on the Infrared-Radio Relation. I. New Constraints on Selection Biases and the Non-Evolution of the Infrared/Radio Properties of Star-Forming and Active Galactic Nucleus Galaxies at Intermediate and High Redshift}",
      journal = {\apjs},
         year = 2010,
        month = feb,
       volume = {186},
       number = {2},
        pages = {341-377},
          doi = {10.1088/0067-0049/186/2/341},
archivePrefix = {arXiv},
       eprint = {1001.1354},
 primaryClass = {astro-ph.CO},
       adsurl = {https://ui.adsabs.harvard.edu/abs/2010ApJS..186..341S}
}

@ARTICLE{sargent2010b,
       author = {{Sargent}, M.~T. and {Schinnerer}, E. and {Murphy}, E. and {Carilli}, C.~L. and {Helou}, G. and {Aussel}, H. and {Le Floc'h}, E. and {Frayer}, D.~T. and {Ilbert}, O. and {Oesch}, P. and {Salvato}, M. and {Smol{\v{c}}i{\'c}}, V. and {Kartaltepe}, J. and {Sanders}, D.~B.},
        title = "{No Evolution in the IR-Radio Relation for IR-luminous Galaxies at z < 2 in the COSMOS Field}",
      journal = {\apjl},
         year = 2010,
        month = may,
       volume = {714},
       number = {2},
        pages = {L190-L195},
          doi = {10.1088/2041-8205/714/2/L190},
archivePrefix = {arXiv},
       eprint = {1003.4271},
 primaryClass = {astro-ph.CO},
       adsurl = {https://ui.adsabs.harvard.edu/abs/2010ApJ...714L.190S}
}

@ARTICLE{condon1992,
       author = {{Condon}, J.~J.},
        title = "{Radio emission from normal galaxies.}",
      journal = {\araa},
         year = 1992,
        month = jan,
       volume = {30},
        pages = {575-611},
          doi = {10.1146/annurev.aa.30.090192.003043},
       adsurl = {https://ui.adsabs.harvard.edu/abs/1992ARA&A..30..575C}
}

@ARTICLE{algera2020_irrc,
       author = {{Algera}, H.~S.~B. and {Smail}, I. and {Dudzevi{\v{c}}i{\={u}}t{\.{e}}}, U. and {Swinbank}, A.~M. and {Stach}, S. and {Hodge}, J.~A. and {Thomson}, A.~P. and {Almaini}, O. and {Arumugam}, V. and {Blain}, A.~W. and {Calistro-Rivera}, G. and {Chapman}, S.~C. and {Chen}, C. -C. and {da Cunha}, E. and {Farrah}, D. and {Leslie}, S. and {Scott}, D. and {van der Vlugt}, D. and {Wardlow}, J.~L. and {van der Werf}, P.},
        title = "{An ALMA Survey of the SCUBA-2 Cosmology Legacy Survey UKIDSS/UDS Field: The Far-infrared/Radio Correlation for High-redshift Dusty Star-forming Galaxies}",
      journal = {\apj},
         year = 2020,
        month = nov,
       volume = {903},
       number = {2},
          eid = {138},
        pages = {138},
          doi = {10.3847/1538-4357/abb77b},
archivePrefix = {arXiv},
       eprint = {2009.06647},
 primaryClass = {astro-ph.GA},
       adsurl = {https://ui.adsabs.harvard.edu/abs/2020ApJ...903..138A}
}

@ARTICLE{murphy2011,
       author = {{Murphy}, E.~J. and {Condon}, J.~J. and {Schinnerer}, E. and {Kennicutt}, R.~C. and {Calzetti}, D. and {Armus}, L. and {Helou}, G. and {Turner}, J.~L. and {Aniano}, G. and {Beir{\~a}o}, P. and {Bolatto}, A.~D. and {Brandl}, B.~R. and {Croxall}, K.~V. and {Dale}, D.~A. and {Donovan Meyer}, J.~L. and {Draine}, B.~T. and {Engelbracht}, C. and {Hunt}, L.~K. and {Hao}, C. -N. and {Koda}, J. and {Roussel}, H. and {Skibba}, R. and {Smith}, J. -D.~T.},
        title = "{Calibrating Extinction-free Star Formation Rate Diagnostics with 33 GHz Free-free Emission in NGC 6946}",
      journal = {\apj},
         year = 2011,
        month = aug,
       volume = {737},
       number = {2},
          eid = {67},
        pages = {67},
          doi = {10.1088/0004-637X/737/2/67},
archivePrefix = {arXiv},
       eprint = {1105.4877},
 primaryClass = {astro-ph.CO},
       adsurl = {https://ui.adsabs.harvard.edu/abs/2011ApJ...737...67M}
}

@ARTICLE{algera2021,
       author = {{Algera}, H.~S.~B. and {Hodge}, J.~A. and {Riechers}, D. and {Murphy}, E.~J. and {Pavesi}, R. and {Aravena}, M. and {Daddi}, E. and {Decarli}, R. and {Dickinson}, M. and {Sargent}, M. and {Sharon}, C.~E. and {Wagg}, J.},
        title = "{COLDz: Deep 34 GHz Continuum Observations and Free-Free Emission in High-redshift Star-forming Galaxies}",
      journal = {\apj},
         year = 2021,
        month = may,
       volume = {912},
       number = {1},
          eid = {73},
        pages = {73},
          doi = {10.3847/1538-4357/abe6a5},
archivePrefix = {arXiv},
       eprint = {2012.08499},
 primaryClass = {astro-ph.GA},
       adsurl = {https://ui.adsabs.harvard.edu/abs/2021ApJ...912...73A}
}

@ARTICLE{chen2024,
       author = {{Chen}, Qingxiang and {Sharon}, Chelsea E. and {Algera}, Hiddo S.~B. and {Baker}, Andrew J. and {Keeton}, Charles R. and {Lutz}, Dieter and {Liu}, Daizhong and {Young}, Anthony J. and {Tagore}, Amitpal S. and {Rivera}, Jesus and {Hicks}, Erin K.~S. and {Allam}, Sahar S. and {Tucker}, Douglas L.},
        title = "{Comparisons between Resolved Star Formation Rate and Gas Tracers in the Strongly Lensed Galaxy SDSS J0901+1814 at Cosmic Noon}",
      journal = {\apj},
         year = 2024,
        month = sep,
       volume = {972},
       number = {2},
          eid = {147},
        pages = {147},
          doi = {10.3847/1538-4357/ad5ceb},
archivePrefix = {arXiv},
       eprint = {2407.01685},
 primaryClass = {astro-ph.GA},
       adsurl = {https://ui.adsabs.harvard.edu/abs/2024ApJ...972..147C}
}

@ARTICLE{fudamoto2021,
       author = {{Fudamoto}, Y. and {Oesch}, P.~A. and {Schouws}, S. and {Stefanon}, M. and {Smit}, R. and {Bouwens}, R.~J. and {Bowler}, R.~A.~A. and {Endsley}, R. and {Gonzalez}, V. and {Inami}, H. and {Labbe}, I. and {Stark}, D. and {Aravena}, M. and {Barrufet}, L. and {da Cunha}, E. and {Dayal}, P. and {Ferrara}, A. and {Graziani}, L. and {Hodge}, J. and {Hutter}, A. and {Li}, Y. and {De Looze}, I. and {Nanayakkara}, T. and {Pallottini}, A. and {Riechers}, D. and {Schneider}, R. and {Ucci}, G. and {van der Werf}, P. and {White}, C.},
        title = "{Normal, dust-obscured galaxies in the epoch of reionization}",
      journal = {\nat},
         year = 2021,
        month = sep,
       volume = {597},
       number = {7877},
        pages = {489-492},
          doi = {10.1038/s41586-021-03846-z},
archivePrefix = {arXiv},
       eprint = {2109.10378},
 primaryClass = {astro-ph.GA},
       adsurl = {https://ui.adsabs.harvard.edu/abs/2021Natur.597..489F}
}

@ARTICLE{casey2018,
       author = {{Casey}, Caitlin M. and {Zavala}, Jorge A. and {Spilker}, Justin and {da Cunha}, Elisabete and {Hodge}, Jacqueline and {Hung}, Chao-Ling and {Staguhn}, Johannes and {Finkelstein}, Steven L. and {Drew}, Patrick},
        title = "{The Brightest Galaxies in the Dark Ages: Galaxies{\textquoteright} Dust Continuum Emission during the Reionization Era}",
      journal = {\apj},
         year = 2018,
        month = jul,
       volume = {862},
       number = {1},
          eid = {77},
        pages = {77},
          doi = {10.3847/1538-4357/aac82d},
archivePrefix = {arXiv},
       eprint = {1805.10301},
 primaryClass = {astro-ph.GA},
       adsurl = {https://ui.adsabs.harvard.edu/abs/2018ApJ...862...77C}
}

@ARTICLE{novak2017,
       author = {{Novak}, M. and {Smol{\v{c}}i{\'c}}, V. and {Delhaize}, J. and {Delvecchio}, I. and {Zamorani}, G. and {Baran}, N. and {Bondi}, M. and {Capak}, P. and {Carilli}, C.~L. and {Ciliegi}, P. and {Civano}, F. and {Ilbert}, O. and {Karim}, A. and {Laigle}, C. and {Le F{\`e}vre}, O. and {Marchesi}, S. and {McCracken}, H. and {Miettinen}, O. and {Salvato}, M. and {Sargent}, M. and {Schinnerer}, E. and {Tasca}, L.},
        title = "{The VLA-COSMOS 3 GHz Large Project: Cosmic star formation history since z   5}",
      journal = {\aap},
         year = 2017,
        month = jun,
       volume = {602},
          eid = {A5},
        pages = {A5},
          doi = {10.1051/0004-6361/201629436},
archivePrefix = {arXiv},
       eprint = {1703.09724},
 primaryClass = {astro-ph.GA},
       adsurl = {https://ui.adsabs.harvard.edu/abs/2017A&A...602A...5N}
}

@ARTICLE{vandervlugt2022,
       author = {{van der Vlugt}, D. and {Hodge}, J.~A. and {Algera}, H.~S.~B. and {Smail}, I. and {Leslie}, S.~K. and {Radcliffe}, J.~F. and {Riechers}, D.~A. and {R{\"o}ttgering}, H.},
        title = "{An Ultra-deep Multiband Very Large Array (VLA) Survey of the Faint Radio Sky (COSMOS-XS): New Constraints on the Cosmic Star Formation History}",
      journal = {\apj},
         year = 2022,
        month = dec,
       volume = {941},
       number = {1},
          eid = {10},
        pages = {10},
          doi = {10.3847/1538-4357/ac99db},
archivePrefix = {arXiv},
       eprint = {2204.04167},
 primaryClass = {astro-ph.GA},
       adsurl = {https://ui.adsabs.harvard.edu/abs/2022ApJ...941...10V}
}

@ARTICLE{bell2003,
       author = {{Bell}, Eric F.},
        title = "{Estimating Star Formation Rates from Infrared and Radio Luminosities: The Origin of the Radio-Infrared Correlation}",
      journal = {\apj},
         year = 2003,
        month = apr,
       volume = {586},
       number = {2},
        pages = {794-813},
          doi = {10.1086/367829},
archivePrefix = {arXiv},
       eprint = {astro-ph/0212121},
 primaryClass = {astro-ph},
       adsurl = {https://ui.adsabs.harvard.edu/abs/2003ApJ...586..794B}
}

@ARTICLE{murphy2017,
       author = {{Murphy}, Eric J. and {Momjian}, Emmanuel and {Condon}, James J. and {Chary}, Ranga-Ram and {Dickinson}, Mark and {Inami}, Hanae and {Taylor}, Andrew R. and {Weiner}, Benjamin J.},
        title = "{The GOODS-N Jansky VLA 10 GHz Pilot Survey: Sizes of Star-forming {\ensuremath{\mu}}JY Radio Sources}",
      journal = {\apj},
         year = 2017,
        month = apr,
       volume = {839},
       number = {1},
          eid = {35},
        pages = {35},
          doi = {10.3847/1538-4357/aa62fd},
archivePrefix = {arXiv},
       eprint = {1702.06963},
 primaryClass = {astro-ph.GA},
       adsurl = {https://ui.adsabs.harvard.edu/abs/2017ApJ...839...35M}
}

@ARTICLE{jimenez-andrade2024,
       author = {{Jim{\'e}nez-Andrade}, Eric F. and {Murphy}, Eric J. and {Momjian}, Emmanuel and {Condon}, James J. and {Chary}, Ranga-Ram and {Taylor}, Russ and {Dickinson}, Mark},
        title = "{A Census of the Deep Radio Sky with the VLA. I. 10 GHz Survey of the GOODS-N Field}",
      journal = {\apj},
         year = 2024,
        month = sep,
       volume = {972},
       number = {1},
          eid = {89},
        pages = {89},
          doi = {10.3847/1538-4357/ad5b5c},
archivePrefix = {arXiv},
       eprint = {2406.13801},
 primaryClass = {astro-ph.GA},
       adsurl = {https://ui.adsabs.harvard.edu/abs/2024ApJ...972...89J}
}

@ARTICLE{bonaldi2019,
       author = {{Bonaldi}, Anna and {Bonato}, Matteo and {Galluzzi}, Vincenzo and {Harrison}, Ian and {Massardi}, Marcella and {Kay}, Scott and {De Zotti}, Gianfranco and {Brown}, Michael L.},
        title = "{The Tiered Radio Extragalactic Continuum Simulation (T-RECS)}",
      journal = {\mnras},
         year = 2019,
        month = jan,
       volume = {482},
       number = {1},
        pages = {2-19},
          doi = {10.1093/mnras/sty2603},
archivePrefix = {arXiv},
       eprint = {1805.05222},
 primaryClass = {astro-ph.GA},
       adsurl = {https://ui.adsabs.harvard.edu/abs/2019MNRAS.482....2B}
}

@ARTICLE{peluso2025,
       author = {{Peluso}, Giorgia and {Delvecchio}, Ivan and {Radcliffe}, Jack and {Daddi}, Emanuele and {Deane}, Roger and {Jarvis}, Matt and {Zamorani}, Giovanni and {Prandoni}, Isabella and {Gitti}, Myriam and {Spingola}, Cristiana and {Ubertosi}, Francesco and {Sargent}, Mark and {Smolcic}, Vernesa and {Wang}, Wuji and {Delhaize}, Jacinta and {Jin}, Shuowen and {Deller}, Adam},
        title = "{Investigating the influence of radio-faint AGN activity on the infrared-radio correlation of massive galaxies}",
      journal = {arXiv e-prints},
         year = 2025,
        month = sep,
          eid = {arXiv:2509.17536},
        pages = {arXiv:2509.17536},
          doi = {10.48550/arXiv.2509.17536},
archivePrefix = {arXiv},
       eprint = {2509.17536},
 primaryClass = {astro-ph.GA},
       adsurl = {https://ui.adsabs.harvard.edu/abs/2025arXiv250917536P}
}

@ARTICLE{algera2020_xs,
       author = {{Algera}, H.~S.~B. and {van der Vlugt}, D. and {Hodge}, J.~A. and {Smail}, I.~R. and {Novak}, M. and {Radcliffe}, J.~F. and {Riechers}, D.~A. and {R{\"o}ttgering}, H. and {Smol{\v{c}}i{\'c}}, V. and {Walter}, F.},
        title = "{A Multiwavelength Analysis of the Faint Radio Sky (COSMOS-XS): the Nature of the Ultra-faint Radio Population}",
      journal = {\apj},
         year = 2020,
        month = nov,
       volume = {903},
       number = {2},
          eid = {139},
        pages = {139},
          doi = {10.3847/1538-4357/abb77a},
archivePrefix = {arXiv},
       eprint = {2009.13531},
 primaryClass = {astro-ph.GA},
       adsurl = {https://ui.adsabs.harvard.edu/abs/2020ApJ...903..139A}
}

@ARTICLE{smolcic2017_release,
       author = {{Smol{\v{c}}i{\'c}}, V. and {Novak}, M. and {Bondi}, M. and {Ciliegi}, P. and {Mooley}, K.~P. and {Schinnerer}, E. and {Zamorani}, G. and {Navarrete}, F. and {Bourke}, S. and {Karim}, A. and {Vardoulaki}, E. and {Leslie}, S. and {Delhaize}, J. and {Carilli}, C.~L. and {Myers}, S.~T. and {Baran}, N. and {Delvecchio}, I. and {Miettinen}, O. and {Banfield}, J. and {Balokovi{\'c}}, M. and {Bertoldi}, F. and {Capak}, P. and {Frail}, D.~A. and {Hallinan}, G. and {Hao}, H. and {Herrera Ruiz}, N. and {Horesh}, A. and {Ilbert}, O. and {Intema}, H. and {Jeli{\'c}}, V. and {Kl{\"o}ckner}, H. -R. and {Krpan}, J. and {Kulkarni}, S.~R. and {McCracken}, H. and {Laigle}, C. and {Middleberg}, E. and {Murphy}, E.~J. and {Sargent}, M. and {Scoville}, N.~Z. and {Sheth}, K.},
        title = "{The VLA-COSMOS 3 GHz Large Project: Continuum data and source catalog release}",
      journal = {\aap},
         year = 2017,
        month = jun,
       volume = {602},
          eid = {A1},
        pages = {A1},
          doi = {10.1051/0004-6361/201628704},
archivePrefix = {arXiv},
       eprint = {1703.09713},
 primaryClass = {astro-ph.GA},
       adsurl = {https://ui.adsabs.harvard.edu/abs/2017A&A...602A...1S}
}

@ARTICLE{smolcic2017_multiwave,
       author = {{Smol{\v{c}}i{\'c}}, V. and {Delvecchio}, I. and {Zamorani}, G. and {Baran}, N. and {Novak}, M. and {Delhaize}, J. and {Schinnerer}, E. and {Berta}, S. and {Bondi}, M. and {Ciliegi}, P. and {Capak}, P. and {Civano}, F. and {Karim}, A. and {Le Fevre}, O. and {Ilbert}, O. and {Laigle}, C. and {Marchesi}, S. and {McCracken}, H.~J. and {Tasca}, L. and {Salvato}, M. and {Vardoulaki}, E.},
        title = "{The VLA-COSMOS 3 GHz Large Project: Multiwavelength counterparts and the composition of the faint radio population}",
      journal = {\aap},
         year = 2017,
        month = jun,
       volume = {602},
          eid = {A2},
        pages = {A2},
          doi = {10.1051/0004-6361/201630223},
archivePrefix = {arXiv},
       eprint = {1703.09719},
 primaryClass = {astro-ph.GA},
       adsurl = {https://ui.adsabs.harvard.edu/abs/2017A&A...602A...2S}
}

@ARTICLE{delvecchio2017,
       author = {{Delvecchio}, I. and {Smol{\v{c}}i{\'c}}, V. and {Zamorani}, G. and {Lagos}, C. Del P. and {Berta}, S. and {Delhaize}, J. and {Baran}, N. and {Alexander}, D.~M. and {Rosario}, D.~J. and {Gonzalez-Perez}, V. and {Ilbert}, O. and {Lacey}, C.~G. and {Le F{\`e}vre}, O. and {Miettinen}, O. and {Aravena}, M. and {Bondi}, M. and {Carilli}, C. and {Ciliegi}, P. and {Mooley}, K. and {Novak}, M. and {Schinnerer}, E. and {Capak}, P. and {Civano}, F. and {Fanidakis}, N. and {Herrera Ruiz}, N. and {Karim}, A. and {Laigle}, C. and {Marchesi}, S. and {McCracken}, H.~J. and {Middleberg}, E. and {Salvato}, M. and {Tasca}, L.},
        title = "{The VLA-COSMOS 3 GHz Large Project: AGN and host-galaxy properties out to z {\ensuremath{\lesssim}} 6}",
      journal = {\aap},
         year = 2017,
        month = jun,
       volume = {602},
          eid = {A3},
        pages = {A3},
          doi = {10.1051/0004-6361/201629367},
archivePrefix = {arXiv},
       eprint = {1703.09720},
 primaryClass = {astro-ph.GA},
       adsurl = {https://ui.adsabs.harvard.edu/abs/2017A&A...602A...3D}
}

@ARTICLE{lyu2022,
       author = {{Lyu}, Jianwei and {Alberts}, Stacey and {Rieke}, George H. and {Rujopakarn}, Wiphu},
        title = "{AGN Selection and Demographics in GOODS-S/HUDF from X-Ray to Radio}",
      journal = {\apj},
         year = 2022,
        month = dec,
       volume = {941},
       number = {2},
          eid = {191},
        pages = {191},
          doi = {10.3847/1538-4357/ac9e5d},
archivePrefix = {arXiv},
       eprint = {2209.06219},
 primaryClass = {astro-ph.GA},
       adsurl = {https://ui.adsabs.harvard.edu/abs/2022ApJ...941..191L}
}

@ARTICLE{an2021,
       author = {{An}, Fangxia and {Vaccari}, M. and {Smail}, Ian and {Jarvis}, M.~J. and {Whittam}, I.~H. and {Hale}, C.~L. and {Jin}, S. and {Collier}, J.~D. and {Daddi}, E. and {Delhaize}, J. and {Frank}, B. and {Murphy}, E.~J. and {Prescott}, M. and {Sekhar}, S. and {Taylor}, A.~R. and {Ao}, Y. and {Knowles}, K. and {Marchetti}, L. and {Randriamampandry}, S.~M. and {Randriamanakoto}, Z.},
        title = "{Radio spectral properties of star-forming galaxies in the MIGHTEE-COSMOS field and their impact on the far-infrared-radio correlation}",
      journal = {\mnras},
         year = 2021,
        month = oct,
       volume = {507},
       number = {2},
        pages = {2643-2658},
          doi = {10.1093/mnras/stab2290},
archivePrefix = {arXiv},
       eprint = {2108.02778},
 primaryClass = {astro-ph.GA},
       adsurl = {https://ui.adsabs.harvard.edu/abs/2021MNRAS.507.2643A}
}

@ARTICLE{calistro-rivera2017,
       author = {{Calistro Rivera}, G. and {Williams}, W.~L. and {Hardcastle}, M.~J. and {Duncan}, K. and {R{\"o}ttgering}, H.~J.~A. and {Best}, P.~N. and {Br{\"u}ggen}, M. and {Chy{\.z}y}, K.~T. and {Conselice}, C.~J. and {de Gasperin}, F. and {Engels}, D. and {G{\"u}rkan}, G. and {Intema}, H.~T. and {Jarvis}, M.~J. and {Mahony}, E.~K. and {Miley}, G.~K. and {Morabito}, L.~K. and {Prandoni}, I. and {Sabater}, J. and {Smith}, D.~J.~B. and {Tasse}, C. and {van der Werf}, P.~P. and {White}, G.~J.},
        title = "{The LOFAR window on star-forming galaxies and AGNs - curved radio SEDs and IR-radio correlation at 0<z<2.5}",
      journal = {\mnras},
         year = 2017,
        month = aug,
       volume = {469},
       number = {3},
        pages = {3468-3488},
          doi = {10.1093/mnras/stx1040},
archivePrefix = {arXiv},
       eprint = {1704.06268},
 primaryClass = {astro-ph.GA},
       adsurl = {https://ui.adsabs.harvard.edu/abs/2017MNRAS.469.3468C}
}

@ARTICLE{tisanic2019,
       author = {{Tisani{\'c}}, K. and {Smol{\v{c}}i{\'c}}, V. and {Delhaize}, J. and {Novak}, M. and {Intema}, H. and {Delvecchio}, I. and {Schinnerer}, E. and {Zamorani}, G. and {Bondi}, M. and {Vardoulaki}, E.},
        title = "{The VLA-COSMOS 3 GHz Large Project: Average radio spectral energy distribution of highly star-forming galaxies}",
      journal = {\aap},
         year = 2019,
        month = jan,
       volume = {621},
          eid = {A139},
        pages = {A139},
          doi = {10.1051/0004-6361/201834002},
archivePrefix = {arXiv},
       eprint = {1812.03392},
 primaryClass = {astro-ph.GA},
       adsurl = {https://ui.adsabs.harvard.edu/abs/2019A&A...621A.139T}
}

@ARTICLE{algera2022,
       author = {{Algera}, Hiddo S.~B. and {Hodge}, Jacqueline A. and {Riechers}, Dominik A. and {Leslie}, Sarah K. and {Smail}, Ian and {Aravena}, Manuel and {Cunha}, Elisabete da and {Daddi}, Emanuele and {Decarli}, Roberto and {Dickinson}, Mark and {Gim}, Hansung B. and {Guaita}, Lucia and {Magnelli}, Benjamin and {Murphy}, Eric J. and {Pavesi}, Riccardo and {Sargent}, Mark T. and {Sharon}, Chelsea E. and {Wagg}, Jeff and {Walter}, Fabian and {Yun}, Min},
        title = "{COLDz: Probing Cosmic Star Formation With Radio Free-Free Emission}",
      journal = {\apj},
         year = 2022,
        month = jan,
       volume = {924},
       number = {2},
          eid = {76},
        pages = {76},
          doi = {10.3847/1538-4357/ac34f5},
archivePrefix = {arXiv},
       eprint = {2111.01153},
 primaryClass = {astro-ph.GA},
       adsurl = {https://ui.adsabs.harvard.edu/abs/2022ApJ...924...76A}
}

@ARTICLE{klein2018,
       author = {{Klein}, U. and {Lisenfeld}, U. and {Verley}, S.},
        title = "{Radio synchrotron spectra of star-forming galaxies}",
      journal = {\aap},
         year = 2018,
        month = mar,
       volume = {611},
          eid = {A55},
        pages = {A55},
          doi = {10.1051/0004-6361/201731673},
archivePrefix = {arXiv},
       eprint = {1710.03149},
 primaryClass = {astro-ph.GA},
       adsurl = {https://ui.adsabs.harvard.edu/abs/2018A&A...611A..55K}
}

@ARTICLE{niklas1997,
       author = {{Niklas}, S. and {Klein}, U. and {Wielebinski}, R.},
        title = "{A radio continuum survey of Shapley-Ames galaxies at {\ensuremath{\lambda}} 2.8cm. II. Separation of thermal and non-thermal radio emission.}",
      journal = {\aap},
         year = 1997,
        month = jun,
       volume = {322},
        pages = {19-28},
       adsurl = {https://ui.adsabs.harvard.edu/abs/1997A&A...322...19N}
}

@ARTICLE{tabatabaei2017,
       author = {{Tabatabaei}, F.~S. and {Schinnerer}, E. and {Krause}, M. and {Dumas}, G. and {Meidt}, S. and {Damas-Segovia}, A. and {Beck}, R. and {Murphy}, E.~J. and {Mulcahy}, D.~D. and {Groves}, B. and {Bolatto}, A. and {Dale}, D. and {Galametz}, M. and {Sandstrom}, K. and {Boquien}, M. and {Calzetti}, D. and {Kennicutt}, R.~C. and {Hunt}, L.~K. and {De Looze}, I. and {Pellegrini}, E.~W.},
        title = "{The Radio Spectral Energy Distribution and Star-formation Rate Calibration in Galaxies}",
      journal = {\apj},
         year = 2017,
        month = feb,
       volume = {836},
       number = {2},
          eid = {185},
        pages = {185},
          doi = {10.3847/1538-4357/836/2/185},
archivePrefix = {arXiv},
       eprint = {1611.01705},
 primaryClass = {astro-ph.GA},
       adsurl = {https://ui.adsabs.harvard.edu/abs/2017ApJ...836..185T}
}

@ARTICLE{thomson2019,
       author = {{Thomson}, A.~P. and {Smail}, Ian and {Swinbank}, A.~M. and {Simpson}, J.~M. and {Arumugam}, V. and {Stach}, S. and {Murphy}, E.~J. and {Rujopakarn}, W. and {Almaini}, O. and {An}, F. and {Blain}, A.~W. and {Chen}, C.~C. and {Cooke}, E.~A. and {Dudzevi{\v{c}}i{\={u}}t{\.{e}}}, U. and {Edge}, A.~C. and {Farrah}, D. and {Gullberg}, B. and {Hartley}, W. and {Ibar}, E. and {Maltby}, D. and {Micha{\l}owski}, M.~J. and {Simpson}, C. and {van der Werf}, P. and {Wardlow}, J.~L.},
        title = "{Radio Spectra and Sizes of Atacama Large Millimeter/submillimeter Array-identified Submillimeter Galaxies: Evidence of Age-related Spectral Curvature and Cosmic-Ray Diffusion?}",
      journal = {\apj},
         year = 2019,
        month = oct,
       volume = {883},
       number = {2},
          eid = {204},
        pages = {204},
          doi = {10.3847/1538-4357/ab32e7},
archivePrefix = {arXiv},
       eprint = {1904.08944},
 primaryClass = {astro-ph.GA},
       adsurl = {https://ui.adsabs.harvard.edu/abs/2019ApJ...883..204T}
}

@ARTICLE{tabatabaei2025,
       author = {{Tabatabaei}, Fatemeh and {Khademi}, Maryam and {Jarvis}, Matt J. and {Taylor}, Russ and {Whittam}, Imogen H. and {An}, Fangxia and {Javadi}, Reihaneh and {Murphy}, Eric J. and {Vaccari}, Mattia},
        title = "{The Radio Spectral Energy Distribution and Star Formation Calibration in MIGHTEE-COSMOS Highly Star-forming Galaxies at 1.5 < z < 3.5}",
      journal = {\apj},
         year = 2025,
        month = aug,
       volume = {989},
       number = {1},
          eid = {44},
        pages = {44},
          doi = {10.3847/1538-4357/ade233},
archivePrefix = {arXiv},
       eprint = {2506.16275},
 primaryClass = {astro-ph.GA},
       adsurl = {https://ui.adsabs.harvard.edu/abs/2025ApJ...989...44T}
}

@ARTICLE{murphy2010,
       author = {{Murphy}, E.~J. and {Helou}, G. and {Condon}, J.~J. and {Schinnerer}, E. and {Turner}, J.~L. and {Beck}, R. and {Mason}, B.~S. and {Chary}, R.-R. and {Armus}, L.},
        title = "{The Detection of Anomalous Dust Emission in the Nearby Galaxy NGC 6946}",
      journal = {\apjl},
         year = 2010,
        month = feb,
       volume = {709},
       number = {2},
        pages = {L108-L113},
          doi = {10.1088/2041-8205/709/2/L108},
archivePrefix = {arXiv},
       eprint = {0912.2731},
 primaryClass = {astro-ph.CO},
       adsurl = {https://ui.adsabs.harvard.edu/abs/2010ApJ...709L.108M}
}

@ARTICLE{murphy2018,
       author = {{Murphy}, E.~J. and {Linden}, S.~T. and {Dong}, D. and {Hensley}, B.~S. and {Momjian}, E. and {Helou}, G. and {Evans}, A.~S.},
        title = "{A New Detection of Extragalactic Anomalous Microwave Emission in a Compact, Optically Faint Region of NGC 4725}",
      journal = {\apj},
         year = 2018,
        month = jul,
       volume = {862},
       number = {1},
          eid = {20},
        pages = {20},
          doi = {10.3847/1538-4357/aac5f5},
archivePrefix = {arXiv},
       eprint = {1805.05965},
 primaryClass = {astro-ph.GA},
       adsurl = {https://ui.adsabs.harvard.edu/abs/2018ApJ...862...20M}
}

@ARTICLE{conway2018,
       author = {{Conway}, J. and {Beswick}, R. and {Bourke}, T. and {Coriat}, M. and {Ferrari}, C. and {Jimenez-Serra}, I. and {Muller}, S. and {Sarent}, M.},
        title = "{SKA1 Beyond 15 GHz: The Science case for Band 6}",
      journal = {SKA Memos},
         year = 2020,
        month = feb,
       volume = {SKA Memos},
       number = {},
          eid = {1},
        pages = {131},
          doi = {},
archivePrefix = {},
       eprint = {},
 primaryClass = {},
       adsurl = {https://www.skao.int/en/science-users/122/relevant-documents}
}

@ARTICLE{condon1997,
       author = {{Condon}, J.~J.},
        title = "{Errors in Elliptical Gaussian Fits}",
      journal = {\pasp},
         year = 1997,
        month = feb,
       volume = {109},
        pages = {166-172},
          doi = {10.1086/133871},
       adsurl = {https://ui.adsabs.harvard.edu/abs/1997PASP..109..166C}
}

@ARTICLE{linden2020,
       author = {{Linden}, S.~T. and {Murphy}, E.~J. and {Dong}, D. and {Momjian}, E. and {Kennicutt}, Jr., R.~C. and {Meier}, D.~S. and {Schinnerer}, E. and {Turner}, J.~L.},
        title = "{The Star Formation in Radio Survey: 3-33 GHz Imaging of Nearby Galaxy Nuclei and Extranuclear Star-forming Regions}",
      journal = {\apjs},
         year = 2020,
        month = jun,
       volume = {248},
       number = {2},
          eid = {25},
        pages = {25},
          doi = {10.3847/1538-4365/ab8a4d},
archivePrefix = {arXiv},
       eprint = {2004.10230},
 primaryClass = {astro-ph.GA},
       adsurl = {https://ui.adsabs.harvard.edu/abs/2020ApJS..248...25L}
}

@ARTICLE{foreman-mackey2013,
       author = {{Foreman-Mackey}, Daniel and {Hogg}, David W. and {Lang}, Dustin and {Goodman}, Jonathan},
        title = "{emcee: The MCMC Hammer}",
      journal = {\pasp},
         year = 2013,
        month = mar,
       volume = {125},
       number = {925},
        pages = {306},
          doi = {10.1086/670067},
archivePrefix = {arXiv},
       eprint = {1202.3665},
 primaryClass = {astro-ph.IM},
       adsurl = {https://ui.adsabs.harvard.edu/abs/2013PASP..125..306F}
}

@ARTICLE{galvin2018,
       author = {{Galvin}, T.~J. and {Seymour}, N. and {Marvil}, J. and {Filipovi{\'c}}, M.~D. and {Tothill}, N.~F.~H. and {McDermid}, R.~M. and {Hurley-Walker}, N. and {Hancock}, P.~J. and {Callingham}, J.~R. and {Cook}, R.~H. and {Norris}, R.~P. and {Bell}, M.~E. and {Dwarakanath}, K.~S. and {For}, B. and {Gaensler}, B.~M. and {Hindson}, L. and {Johnston-Hollitt}, M. and {Kapi{\'n}ska}, A.~D. and {Lenc}, E. and {McKinley}, B. and {Morgan}, J. and {Offringa}, A.~R. and {Procopio}, P. and {Staveley-Smith}, L. and {Wayth}, R.~B. and {Wu}, C. and {Zheng}, Q.},
        title = "{The spectral energy distribution of powerful starburst galaxies - I. Modelling the radio continuum}",
      journal = {\mnras},
         year = 2018,
        month = feb,
       volume = {474},
       number = {1},
        pages = {779-799},
          doi = {10.1093/mnras/stx2613},
archivePrefix = {arXiv},
       eprint = {1710.01967},
 primaryClass = {astro-ph.GA},
       adsurl = {https://ui.adsabs.harvard.edu/abs/2018MNRAS.474..779G}
}

@ARTICLE{jimenez-andrade2019,
       author = {{Jim{\'e}nez-Andrade}, E.~F. and {Magnelli}, B. and {Karim}, A. and {Zamorani}, G. and {Bondi}, M. and {Schinnerer}, E. and {Sargent}, M. and {Romano-D{\'\i}az}, E. and {Novak}, M. and {Lang}, P. and {Bertoldi}, F. and {Vardoulaki}, E. and {Toft}, S. and {Smol{\v{c}}i{\'c}}, V. and {Harrington}, K. and {Leslie}, S. and {Delhaize}, J. and {Liu}, D. and {Karoumpis}, C. and {Kartaltepe}, J. and {Koekemoer}, A.~M.},
        title = "{Radio continuum size evolution of star-forming galaxies over 0.35 < z < 2.25}",
      journal = {\aap},
         year = 2019,
        month = may,
       volume = {625},
          eid = {A114},
        pages = {A114},
          doi = {10.1051/0004-6361/201935178},
archivePrefix = {arXiv},
       eprint = {1903.12217},
 primaryClass = {astro-ph.GA},
       adsurl = {https://ui.adsabs.harvard.edu/abs/2019A&A...625A.114J}
}

@ARTICLE{miettinen2017,
       author = {{Miettinen}, O. and {Novak}, M. and {Smol{\v{c}}i{\'c}}, V. and {Delvecchio}, I. and {Aravena}, M. and {Brisbin}, D. and {Karim}, A. and {Murphy}, E.~J. and {Schinnerer}, E. and {Albrecht}, M. and {Aussel}, H. and {Bertoldi}, F. and {Capak}, P.~L. and {Casey}, C.~M. and {Civano}, F. and {Hayward}, C.~C. and {Herrera Ruiz}, N. and {Ilbert}, O. and {Jiang}, C. and {Laigle}, C. and {Le F{\`e}vre}, O. and {Magnelli}, B. and {Marchesi}, S. and {McCracken}, H.~J. and {Middelberg}, E. and {Mu{\~n}oz Arancibia}, A.~M. and {Navarrete}, F. and {Padilla}, N.~D. and {Riechers}, D.~A. and {Salvato}, M. and {Scott}, K.~S. and {Sheth}, K. and {Tasca}, L.~A.~M. and {Bondi}, M. and {Zamorani}, G.},
        title = "{An ALMA survey of submillimetre galaxies in the COSMOS field: The extent of the radio-emitting region revealed by 3 GHz imaging with the Very Large Array}",
      journal = {\aap},
         year = 2017,
        month = jun,
       volume = {602},
          eid = {A54},
        pages = {A54},
          doi = {10.1051/0004-6361/201730443},
archivePrefix = {arXiv},
       eprint = {1702.07527},
 primaryClass = {astro-ph.GA},
       adsurl = {https://ui.adsabs.harvard.edu/abs/2017A&A...602A..54M}
}

@ARTICLE{bouwens2015,
       author = {{Bouwens}, R.~J. and {Illingworth}, G.~D. and {Oesch}, P.~A. and {Trenti}, M. and {Labb{\'e}}, I. and {Bradley}, L. and {Carollo}, M. and {van Dokkum}, P.~G. and {Gonzalez}, V. and {Holwerda}, B. and {Franx}, M. and {Spitler}, L. and {Smit}, R. and {Magee}, D.},
        title = "{UV Luminosity Functions at Redshifts z {\ensuremath{\sim}} 4 to z {\ensuremath{\sim}} 10: 10,000 Galaxies from HST Legacy Fields}",
      journal = {\apj},
         year = 2015,
        month = apr,
       volume = {803},
       number = {1},
          eid = {34},
        pages = {34},
          doi = {10.1088/0004-637X/803/1/34},
archivePrefix = {arXiv},
       eprint = {1403.4295},
 primaryClass = {astro-ph.CO},
       adsurl = {https://ui.adsabs.harvard.edu/abs/2015ApJ...803...34B}
}

@ARTICLE{oesch2018,
       author = {{Oesch}, P.~A. and {Bouwens}, R.~J. and {Illingworth}, G.~D. and {Labb{\'e}}, I. and {Stefanon}, M.},
        title = "{The Dearth of z {\ensuremath{\sim}} 10 Galaxies in All HST Legacy Fields{\textemdash}The Rapid Evolution of the Galaxy Population in the First 500 Myr}",
      journal = {\apj},
         year = 2018,
        month = mar,
       volume = {855},
       number = {2},
          eid = {105},
        pages = {105},
          doi = {10.3847/1538-4357/aab03f},
archivePrefix = {arXiv},
       eprint = {1710.11131},
 primaryClass = {astro-ph.GA},
       adsurl = {https://ui.adsabs.harvard.edu/abs/2018ApJ...855..105O}
}

@ARTICLE{mclure2013,
       author = {{McLure}, R.~J. and {Dunlop}, J.~S. and {Bowler}, R.~A.~A. and {Curtis-Lake}, E. and {Schenker}, M. and {Ellis}, R.~S. and {Robertson}, B.~E. and {Koekemoer}, A.~M. and {Rogers}, A.~B. and {Ono}, Y. and {Ouchi}, M. and {Charlot}, S. and {Wild}, V. and {Stark}, D.~P. and {Furlanetto}, S.~R. and {Cirasuolo}, M. and {Targett}, T.~A.},
        title = "{A new multifield determination of the galaxy luminosity function at z = 7-9 incorporating the 2012 Hubble Ultra-Deep Field imaging}",
      journal = {\mnras},
         year = 2013,
        month = jul,
       volume = {432},
       number = {4},
        pages = {2696-2716},
          doi = {10.1093/mnras/stt627},
archivePrefix = {arXiv},
       eprint = {1212.5222},
 primaryClass = {astro-ph.CO},
       adsurl = {https://ui.adsabs.harvard.edu/abs/2013MNRAS.432.2696M}
}

@ARTICLE{finkelstein2023,
       author = {{Finkelstein}, Steven L. and {Bagley}, Micaela B. and {Ferguson}, Henry C. and {Wilkins}, Stephen M. and {Kartaltepe}, Jeyhan S. and {Papovich}, Casey and {Yung}, L.~Y. Aaron and {Arrabal Haro}, Pablo and {Behroozi}, Peter and {Dickinson}, Mark and {Kocevski}, Dale D. and {Koekemoer}, Anton M. and {Larson}, Rebecca L. and {Le Bail}, Aur{\'e}lien and {Morales}, Alexa M. and {P{\'e}rez-Gonz{\'a}lez}, Pablo G. and {Burgarella}, Denis and {Dav{\'e}}, Romeel and {Hirschmann}, Michaela and {Somerville}, Rachel S. and {Wuyts}, Stijn and {Bromm}, Volker and {Casey}, Caitlin M. and {Fontana}, Adriano and {Fujimoto}, Seiji and {Gardner}, Jonathan P. and {Giavalisco}, Mauro and {Grazian}, Andrea and {Grogin}, Norman A. and {Hathi}, Nimish P. and {Hutchison}, Taylor A. and {Jha}, Saurabh W. and {Jogee}, Shardha and {Kewley}, Lisa J. and {Kirkpatrick}, Allison and {Long}, Arianna S. and {Lotz}, Jennifer M. and {Pentericci}, Laura and {Pierel}, Justin D.~R. and {Pirzkal}, Nor and {Ravindranath}, Swara and {Ryan}, Russell E. and {Trump}, Jonathan R. and {Yang}, Guang and {Bhatawdekar}, Rachana and {Bisigello}, Laura and {Buat}, V{\'e}ronique and {Calabr{\`o}}, Antonello and {Castellano}, Marco and {Cleri}, Nikko J. and {Cooper}, M.~C. and {Croton}, Darren and {Daddi}, Emanuele and {Dekel}, Avishai and {Elbaz}, David and {Franco}, Maximilien and {Gawiser}, Eric and {Holwerda}, Benne W. and {Huertas-Company}, Marc and {Jaskot}, Anne E. and {Leung}, Gene C.~K. and {Lucas}, Ray A. and {Mobasher}, Bahram and {Pandya}, Viraj and {Tacchella}, Sandro and {Weiner}, Benjamin J. and {Zavala}, Jorge A.},
        title = "{CEERS Key Paper. I. An Early Look into the First 500 Myr of Galaxy Formation with JWST}",
      journal = {\apjl},
         year = 2023,
        month = mar,
       volume = {946},
       number = {1},
          eid = {L13},
        pages = {L13},
          doi = {10.3847/2041-8213/acade4},
archivePrefix = {arXiv},
       eprint = {2211.05792},
 primaryClass = {astro-ph.GA},
       adsurl = {https://ui.adsabs.harvard.edu/abs/2023ApJ...946L..13F}
}

@ARTICLE{leslie2020,
       author = {{Leslie}, Sarah K. and {Schinnerer}, Eva and {Liu}, Daizhong and {Magnelli}, Benjamin and {Algera}, Hiddo and {Karim}, Alexander and {Davidzon}, Iary and {Gozaliasl}, Ghassem and {Jim{\'e}nez-Andrade}, Eric F. and {Lang}, Philipp and {Sargent}, Mark T. and {Novak}, Mladen and {Groves}, Brent and {Smol{\v{c}}i{\'c}}, Vernesa and {Zamorani}, Giovanni and {Vaccari}, Mattia and {Battisti}, Andrew and {Vardoulaki}, Eleni and {Peng}, Yingjie and {Kartaltepe}, Jeyhan},
        title = "{The VLA-COSMOS 3 GHz Large Project: Evolution of Specific Star Formation Rates out to z {\ensuremath{\sim}} 5}",
      journal = {\apj},
         year = 2020,
        month = aug,
       volume = {899},
       number = {1},
          eid = {58},
        pages = {58},
          doi = {10.3847/1538-4357/aba044},
archivePrefix = {arXiv},
       eprint = {2006.13937},
 primaryClass = {astro-ph.GA},
       adsurl = {https://ui.adsabs.harvard.edu/abs/2020ApJ...899...58L}
}

@ARTICLE{thomson2012,
       author = {{Thomson}, A.~P. and {Ivison}, R.~J. and {Smail}, Ian and {Swinbank}, A.~M. and {Weiss}, A. and {Kneib}, J.-P. and {Papadopoulos}, P.~P. and {Baker}, A.~J. and {Sharon}, C.~E. and {van Moorsel}, G.~A.},
        title = "{VLA imaging of $^{12}$CO J = 1-0 and free-free emission in lensed submillimetre galaxies}",
      journal = {\mnras},
         year = 2012,
        month = sep,
       volume = {425},
       number = {3},
        pages = {2203-2211},
          doi = {10.1111/j.1365-2966.2012.21584.x},
archivePrefix = {arXiv},
       eprint = {1207.0492},
 primaryClass = {astro-ph.CO},
       adsurl = {https://ui.adsabs.harvard.edu/abs/2012MNRAS.425.2203T}
}

@ARTICLE{huynh2017,
       author = {{Huynh}, Minh T. and {Emonts}, B.~H.~C. and {Kimball}, A.~E. and {Seymour}, N. and {Smail}, Ian and {Swinbank}, A.~M. and {Brandt}, W.~N. and {Casey}, C.~M. and {Chapman}, S.~C. and {Dannerbauer}, H. and {Hodge}, J.~A. and {Ivison}, R.~J. and {Schinnerer}, E. and {Thomson}, A.~P. and {van der Werf}, P. and {Wardlow}, J.~L.},
        title = "{The AT-LESS CO(1-0) survey of submillimetre galaxies in the Extended Chandra Deep Field South: First results on cold molecular gas in galaxies at z {\ensuremath{\sim}} 2}",
      journal = {\mnras},
         year = 2017,
        month = may,
       volume = {467},
       number = {1},
        pages = {1222-1230},
          doi = {10.1093/mnras/stx156},
archivePrefix = {arXiv},
       eprint = {1701.05698},
 primaryClass = {astro-ph.GA},
       adsurl = {https://ui.adsabs.harvard.edu/abs/2017MNRAS.467.1222H}
}

@INPROCEEDINGS{prandoni_seymour2015,
       author = {{Prandoni}, I. and {Seymour}, N.},
        title = "{Revealing the Physics and Evolution of Galaxies and Galaxy Clusters with SKA Continuum Surveys}",
    booktitle = {Advancing Astrophysics with the Square Kilometre Array (AASKA14)},
         year = 2015,
        month = apr,
          eid = {67},
        pages = {67},
          doi = {10.22323/1.215.0067},
archivePrefix = {arXiv},
       eprint = {1412.6512},
 primaryClass = {astro-ph.IM},
       adsurl = {https://ui.adsabs.harvard.edu/abs/2015aska.confE..67P}
}

@incollection{FangxiaAn01.2026.SKA, author = {Fangxia X. An and author2 and author3 and author4 and author5},title = {},year = {2026},publisher = {},note = {arXiv search: Report number AASKAII/FangxiaAn01},booktitle = {Advancing Astrophysics with the SKA -- II (AASKAII)}}
